\documentclass{ieeetj}

\usepackage{amsmath,amsfonts,amsfonts,amsthm}
\usepackage{algorithmic}
\usepackage{algorithm}
\usepackage{array}
\usepackage[caption=false,font=normalsize,labelfont=sf,textfont=sf]{subfig}
\usepackage{textcomp}
\usepackage{stfloats}
\usepackage{url}
\usepackage{verbatim}
\usepackage{graphicx,color}
\usepackage{cite}
\usepackage{xcolor}
\usepackage{optidef}
\usepackage{dsfont}
\usepackage{hyperref}

\newtheorem{theorem}{Theorem}

\newtheorem{definition}{Definition}
\hypersetup{hidelinks=true}

\def\BibTeX{{\rm B\kern-.05em{\sc i\kern-.025em b}\kern-.08em
    T\kern-.1667em\lower.7ex\hbox{E}\kern-.125emX}}
\AtBeginDocument{\definecolor{tmlcncolor}{cmyk}{0.93,0.59,0.15,0.02}\definecolor{NavyBlue}{RGB}{0,86,125}}

\def\OJlogo{\vspace{-4pt}$<$Society logo(s) and publication title will appear here.$>$}
\def\seclogo{\vspace{10pt}$<$Society logo(s) and publication title will appear here.$>$}

\def\authorrefmark#1{\ensuremath{^{\textbf{#1}}}}

\begin{document}
\receiveddate{02 May, 2025}
\reviseddate{XX Month, XXXX}
\accepteddate{XX Month, XXXX}
\publisheddate{XX Month, XXXX}
\currentdate{XX Month, XXXX}
\doiinfo{XXXX.2022.1234567}

\markboth{Journal of Selected Topics in Electromagnetics, Antennas and Propagation,~Vol.~XX, No.~XX, May~2025}%
{Pallaprolu \MakeLowercase{\textit{et al.}}: Embracing Diffraction: A Paradigm Shift in Wireless Sensing and Communication}

\title{Embracing Diffraction: \\ A Paradigm Shift in Wireless Sensing and Communication}

\author{Anurag Pallaprolu\authorrefmark{1},~\IEEEmembership{Student Member,~IEEE,} Winston Hurst\authorrefmark{1},~\IEEEmembership{Student Member,~IEEE,}
and Yasamin Mostofi\authorrefmark{1},~\IEEEmembership{Fellow,~IEEE}}
\affil{Department of Electrical and Computer Engineering, University of California, Santa Barbara, CA 93106 USA}
\corresp{Corresponding author: Anurag Pallaprolu (email: apallaprolu@ucsb.edu).}
\authornote{This work was supported in part by NSF CNS award 2226255, in part by NSF CNS award 2215646, and in part by ONR award N00014-23-1-2715.\vspace*{-30pt}}

\begin{abstract}
Wireless signals are integral to modern society, enabling both communication and increasingly, environmental sensing. While various propagation models exist, ranging from empirical methods to full-wave simulations, the phenomenon of electromagnetic diffraction is often treated as a secondary effect or a correction factor. This paper positions diffraction as a fundamentally important and underutilized mechanism that is rich with information about the physical environment. Specifically, diffraction-inducing elements generate distinct signatures that are rich with information about their underlying properties such as their geometries. We then argue that by understanding and exploiting these relationships, diffraction can be harnessed strategically. We introduce a general optimization framework to formalize this concept, illustrating how diffraction can be leveraged for both inverse problems (sensing scene details such as object geometries from measured fields) and design problems (shaping radio frequency (RF) fields for communication objectives by configuring diffracting elements). Focusing primarily on edge diffraction and Keller's Geometrical Theory of Diffraction (GTD), we discuss specific applications in RF sensing for scene understanding and in communications for RF field programming, drawing upon recent work. Overall, this paper lays out a vision for systematically incorporating diffraction into the design and operation of future wireless systems, paving the way for enhanced sensing capabilities and more robust communication strategies.
\end{abstract}



\maketitle

\section{Introduction}\label{sec:intro}
\IEEEPARstart{W}{ireless} signals permeate our modern world, forming an invisible fabric that connects millions of devices and supports critical societal functions. At the same time, the use of these signals for sensing tasks has become mainstream, as indicated by the inclusion of sensing as a critical aspect of the developing 6G standard~\cite{ITU-R_M.2160-0}. Whether for communication or sensing, intentionally leveraging radio frequency (RF) signals to achieve deliberate objectives relies on understanding and exploiting the underlying physics of RF propagation.

Wireless propagation modeling encompasses a broad spectrum of approaches, from empirical models~\cite{luo2024channel} to full-wave electromagnetic solvers~\cite{ansysHFSS, CSTStudioSuite}, each offering a different balance between physical fidelity and computational cost. On one hand, numerical methods that directly solve Maxwell's equations offer high accuracy but are limited to small environments. On the other hand, simpler conventional models rooted in geometric optics have been successful in tasks like coverage estimation and link budgeting~\cite{zhang2024radio}. The success of this latter approach lies in identifying the fundamental characteristics of RF propagation, such as transmission and reflection, which reveal how key system parameters affect the RF field.

\begin{figure}
    \centering
    \includegraphics[width=\linewidth]{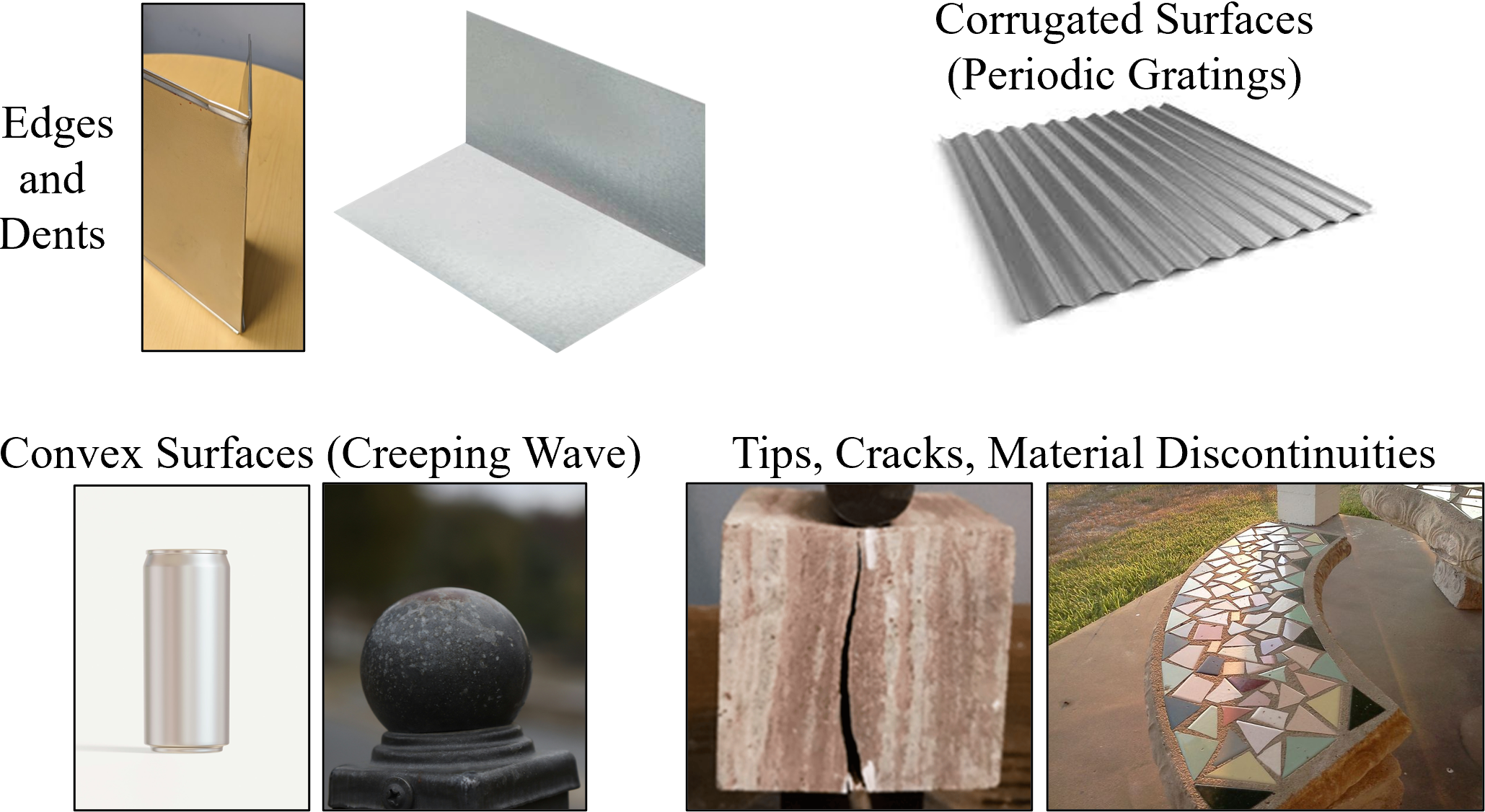}
    \caption{Typical geometries encountered in daily life that cause RF diffraction.}
    \label{fig:sample_objects}
    \vspace{-0.2in}
\end{figure}

In this context, one particularly overlooked phenomenon is \textbf{diffraction}, which describes how electromagnetic waves bend around obstacles and spread into shadow regions. Many daily life objects exhibit geometries that cause diffraction, as shown in Fig.~\ref{fig:sample_objects}. While these effects are captured in high-fidelity ray tracers and electromagnetic field solvers, diffraction is almost always treated as a secondary correction: useful for completeness, but rarely positioned as a strategic tool. Critically, these modeling approaches obscure a key feature of diffraction: its effects are a function of the geometry and material of diffracting objects in a manner that causes the resulting diffraction patterns to carry information-rich signatures of the specific geometric configurations. This has significant implications for both sensing and communication tasks, and in this paper, we lay out our vision of harnessing well-defined relationships between diffracting elements and the resulting RF fields to improve both wireless sensing and communication.

A few specific modes of diffraction have received attention in the literature, most notably Knife-Edge Diffraction (KED)~\cite{bertoni1999radio}, which remains a mainstay due to its analytical simplicity. More recently, some early works have attempted to leverage diffraction for sensing, using, for example, Fresnel zone-based models~\cite{wu2022wifi}. While these works are consistent with our vision, we advocate a broader and more systematic paradigm shift that focuses on diffraction as a controllable and information-rich propagation mechanism.

To this end, we begin with a primer on diffraction in Section~\ref{sec:diffraction_primer}, reviewing key mechanisms, with particular emphasis on edge diffraction. We also introduce a high-level optimization problem that conceptually illustrates how diffraction can be leveraged for sensing or communication purposes. In Section~\ref{sec:sensing}, we explore applications of diffraction in sensing scenarios, drawing from our recent work in~\cite{pallaprolu2022wiffract, pallaprolu2023analysis} to illustrate key concepts, and further discussing other emerging work in this area. Section~\ref{sec:comms} provides insights into how diffraction can be used to improve communications through RF field programming, with our work in \cite{pallaprolu2023beg} serving as an example. We look towards the future in Section~\ref{sec:future}, enumerating challenges and promising research directions, and in Section~\ref{sec:conclusion}, we provide concluding remarks.

\section{A Primer on Electromagnetic Diffraction}\label{sec:diffraction_primer}
Electromagnetic diffraction refers to the subtle ways in which wave propagation deviates from standard geometric optics, particularly in the vicinity of spatial discontinuities such as edges, apertures, and gratings. In his seminal treatise on optics in 1964~\cite{sommerfeld1964optics}, Sommerfeld states that:
\begin{quote}
\textit{“Any deviation of light rays from rectilinear paths which cannot be interpreted as reflection or refraction is called diffraction.”}
\end{quote}
More broadly, diffraction arises from discontinuities in surface geometry, such as abrupt changes in the surface normal, and, in some cases, from discontinuities in material properties. In particular, when the local radius of curvature is no longer large when compared to the wavelength of the incident wave, geometric optics breaks down, and diffraction effects become more pronounced. Diffraction has traditionally been overlooked in the analysis and modeling of communication systems, often treated as a  negligible effect. However, diffraction mechanisms are not only functions of the local properties of the scatterer, e.g., its position, orientation, and curvature, but more importantly can result in unique structured RF signatures. These signatures reflect the specific geometric characteristics of the scatterer, with different local geometries producing distinct RF responses.  

\begin{figure*}
    \centering
    \includegraphics[width=\linewidth]{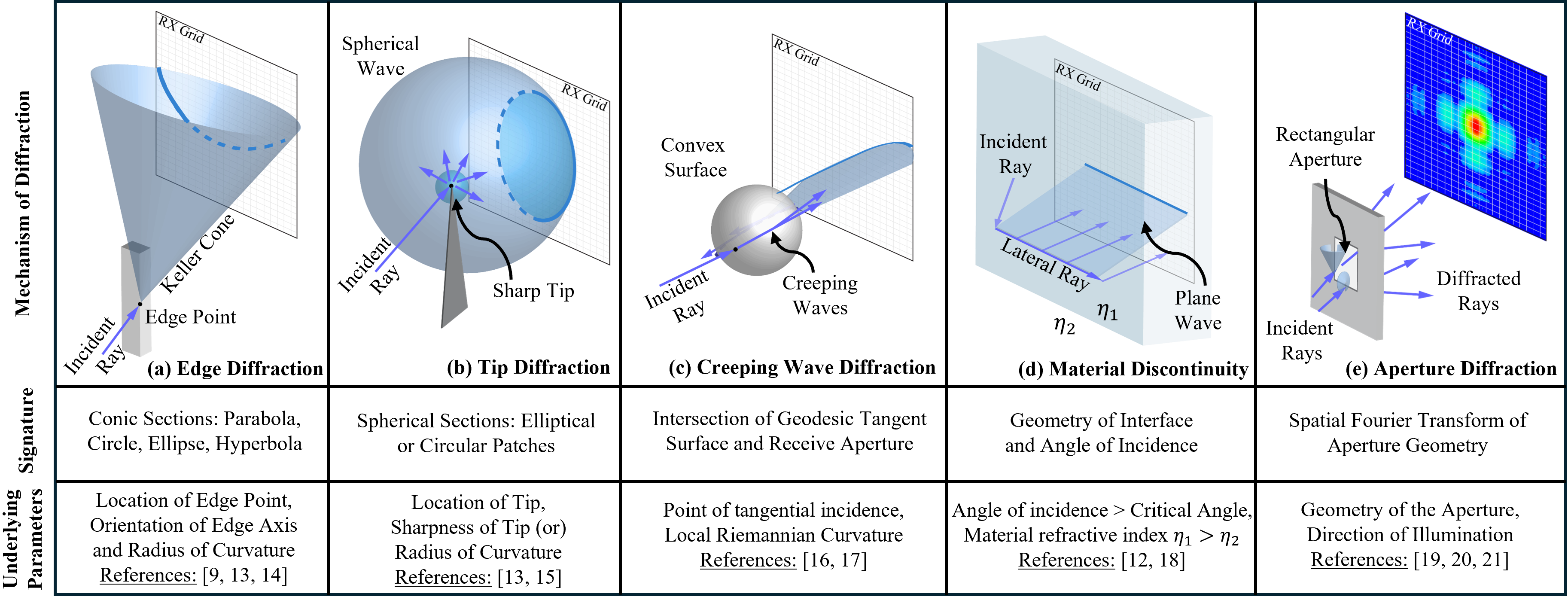}
    \caption{Mechanisms of diffraction, receiver footprints and underlying parameters corresponding to general modes of diffraction encountered in wave propagation.}
    \label{fig:modes_of_diffraction}
    \vspace{-0.2in}
\end{figure*}

This paper then sets forth a new position on the role of diffraction, arguing that it is a rich, yet underutilized mechanism. More specifically, we offer a new perspective for the areas of sensing and communication by showing how geometric degrees of freedom of diffraction-inducing elements can be  harnessed, either to systematically infer the position and surface structure of scatterers for sensing tasks, or to deliberately shape the RF field to achieve communication objectives. In the rest of the section, we begin by introducing canonical examples of structures whose surface geometries can lead to diffraction, with particular emphasis on the characteristic scattering signatures observed on an extended receiver grid. Building on these insights, the subsequent sections then set forth a new foundation for sensing and communication that explicitly leverages the geometric degrees of freedom associated with diffraction-inducing elements.

\subsection{Geometry of Diffracted Ray Propagation}
While diffraction refers to a quite general category of propagation, there are prominent diffraction mechanisms commonly encountered in everyday environments which display distinct and well-defined propagation characteristics~\cite{felsen1994radiation}. Fig.~\ref{fig:modes_of_diffraction} illustrates these mechanisms and representative signatures captured on a vertical receiver grid when illuminated by an RF transmitter. Critically, these signatures depend on certain geometric parameters of the diffracting element, and thus variations in the configuration of the element result in predictable variations in the resulting RF field. We next discuss the mechanisms and their signatures in more detail. In the discussion that follows, we denote the distance between the point of diffraction and the receiver antenna by $r$.

Fig.~\ref{fig:modes_of_diffraction} (a) depicts \textbf{edge diffraction}, whose scattering mechanism is explained by Keller's Geometrical Theory of Diffraction (GTD)\footnote{We note that the Uniform Theory of Diffraction (UTD) is a generalization of GTD that provides uniform field expressions to address caustic regions and avoid singularities. Although this work does not use explicit diffraction coefficients, incorporating UTD can be a promising direction for future extension of this work.}~\cite{keller1962geometrical, kouyoumjian1974uniform}. More specifically, when an incident wave encounters the edge of an object, it generates diffracted rays that radiate along a cone known as the \textbf{Keller cone}. The axis of this cone coincides with the edge, and its angle equals the angle of incidence, with the field strength decaying as $1/\sqrt{r}$. Mathematically, we have the following theorem:

\begin{theorem}[\textbf{Keller's Law of Edge Diffraction}]
\label{thm:keller_law}
Let $\mathbf{p}_t$ be the position of the transmitter, $\mathbf{p}_e$ a point on the illuminated edge with direction $\hat{\mathbf{e}}$, and $\mathbf{p}_r$ a point on the receiver grid. Then, an edge diffracted ray from $\mathbf{p}_e$ arrives at $\mathbf{p}_r$ if \begin{align}\label{eq:conic_intersection} \left\langle\frac{\mathbf{p}_r- \mathbf{p}_e}{\lVert \mathbf{p}_r- \mathbf{p}_e\rVert} + \frac{\mathbf{p}_t- \mathbf{p}_e}{\lVert \mathbf{p}_t- \mathbf{p}_e\rVert},\hat{\mathbf{e}}\right\rangle = 0. \end{align} \end{theorem}

\begin{figure}
    \centering
    \includegraphics[width=\linewidth]{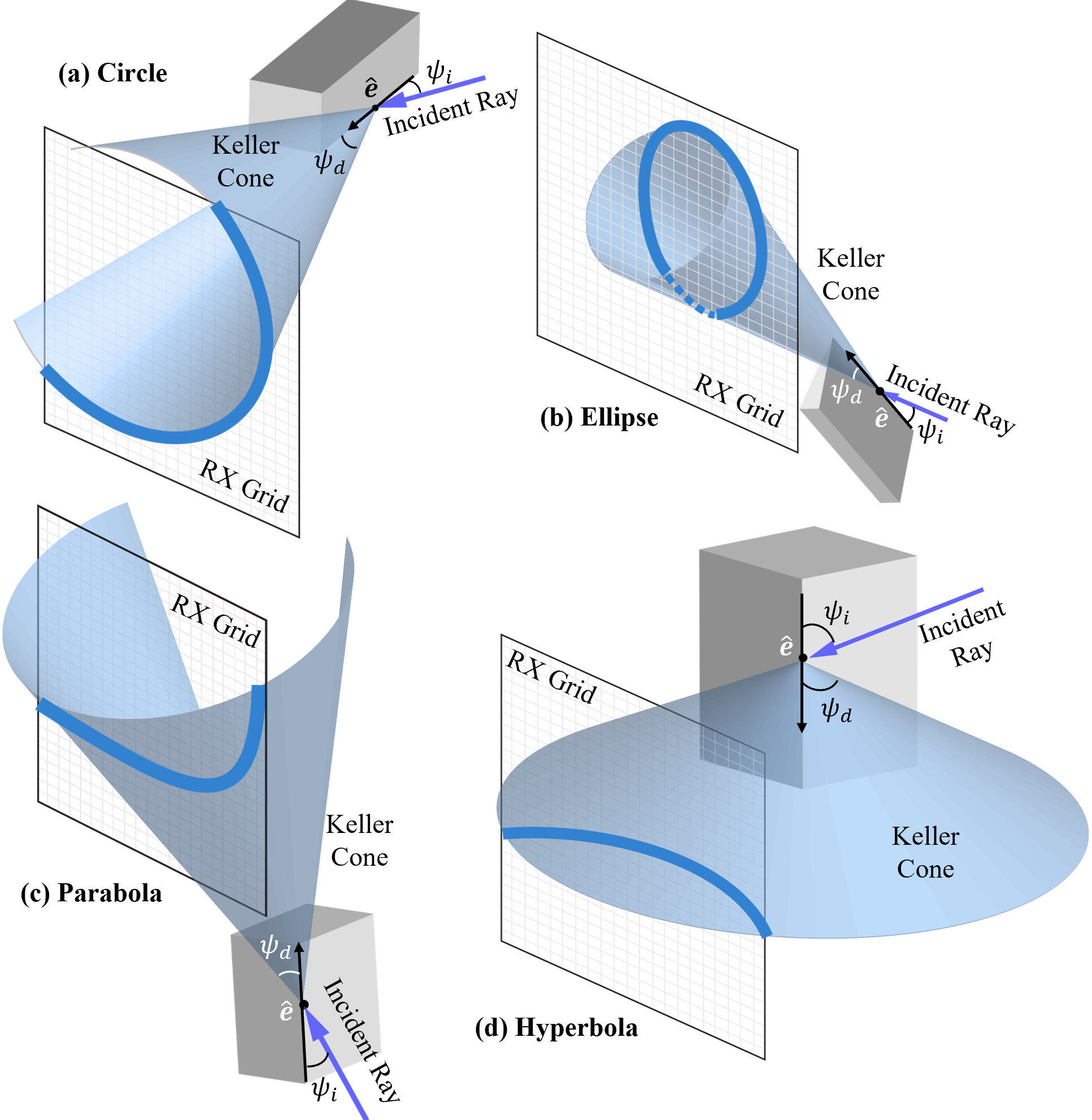}
    \caption{Keller's Law of Edge Diffraction: When a ray is incident on an edge, the diffracted rays lie on a Keller cone, whose axis aligns with the edge direction $\hat{\mathbf{e}}$, and whose angle $\psi_d$ equals the angle of incidence $\psi_i$. Intersection of the Keller cone with a planar receiver grid then yields conic sections (circle, ellipse, parabola, hyperbola) that effectively encode the geometry of the edge.}
    \label{fig:gtd_conics}
\end{figure}

A key geometric consequence of this law is that the intersection of a Keller cone with a planar receiver aperture yields a conic section: a circle, ellipse, parabola, or hyperbola. The shape, location, and underlying parameters of the conic section then depend on the properties of the edge, such as its location and orientation with respect to both the source of the incident radiation and the receiver grid. This dependency effectively imprints a geometric signature onto the RF field, which can be systematically exploited for both inverse problems (e.g., reconstruction and sensing) and design problems (e.g., RF field programming).  Fig.~\ref{fig:gtd_conics} illustrates examples of these four fundamental intersections, further highlighting the resulting structured signatures of edge-induced diffraction, which we shall extensively leverage in Sec.~\ref{sec:sensing} for sensing and and in Sec.~\ref{sec:comms} for RF field programming.

Other diffraction phenomena also produce radiation patterns that depend on the physical properties of the scatterer. Fig.~\ref{fig:modes_of_diffraction} (b) illustrates \textbf{tip diffraction}, wherein a ray interacts with a sharp vertex or point, resulting in a sphere of diffracted rays.
The classical Knife-Edge Diffraction model~\cite{sommerfeld1964optics} provides a first-order approximation for this process, though application of the GTD has been shown to yield more accurate results~\cite{luebbers1984finite}. The resulting field strength decays as $1/r$, consistent with isotropic spreading from a point-like scatterer, and the signature captured on a receiver grid is a circular or an elliptical patch depending on the orientation of the grid. In contrast, when a wave is tangentially incident on a convex surface (Fig.~\ref{fig:modes_of_diffraction} (c)), the scattering process involves \textbf{creeping waves} which are surface-confined rays that travel along geodesics while continually shedding energy tangentially. This phenomenon, central to the diffraction theory also developed by Keller, and extended in later works~\cite{keller1956diffraction, chen2013ray}, is characterized by an exponentially decaying amplitude of the creeping wave and a signature on the receiver grid formed by the intersection of the plane defined by the geodesic tangents with the receive aperture. Fig.~\ref{fig:modes_of_diffraction} (d) showcases diffraction induced by \textbf{material discontinuities}. Specifically, when a wave transitions to a less dense medium at an incidence angle beyond a critical threshold, it undergoes Total Internal Reflection. The associated field gives rise to lateral rays, which travel along the interface and radiate energy back into the denser medium. These diffracted rays maintain their angle of departure equal to the angle of incidence, producing a signature on the receiver grid that is shaped by the geometry of the interface~\cite{felsen1994radiation, brekhovskikh1960waves}. Finally, Fig.~\ref{fig:modes_of_diffraction} (e) illustrates the classical case of \textbf{aperture diffraction}, which was originally treated using Fourier Optics~\cite{goodman1996fourier}. Recent GTD-based models~\cite{balanis2005antenna, zalipaev2021diffraction} represent the field as a superposition of direct rays through the aperture and edge-diffracted rays, with the signature on the receiver grid shaped by the aperture geometry and the resulting Keller cones.

Overall, we see that several diffraction mechanisms produce distinct signatures, with varying degrees of structure and richness, parameterized by the local geometric properties of the diffracting element. In the next subsection, we capitalize on this geometric intuition to formulate a general optimization framework. This framework seeks to exploit the structured signatures of diffraction-inducing elements to estimate scatterer geometry in sensing tasks or, conversely, to program the scattered field for performance enhancement in communication scenarios.

\subsection{Harnessing Diffraction: A General Optimization Problem}
\label{subsec:general_opt}
\begin{figure}
    \centering
    \includegraphics[width=\linewidth]{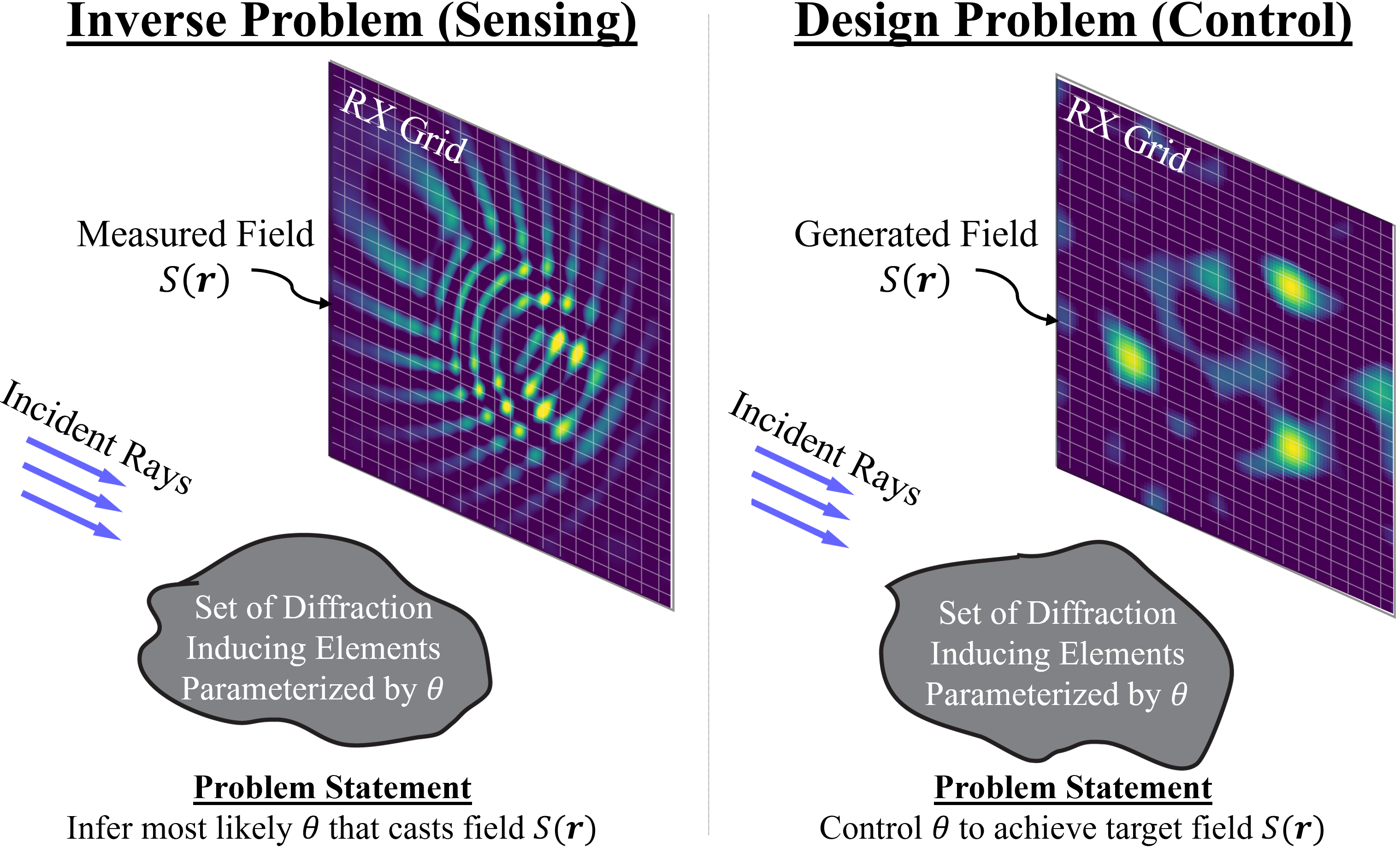}
    \caption{A high-level illustration of the problems considered in this paper. The set of diffraction-inducing elements, shown in gray and parameterized by $\theta$, results in a field $S_\theta(\mathbf{r})$ when illuminated by RF radiation. We seek the configuration, $\theta^\star\in\Theta$ that best matches a target field, $S(\mathbf{r})$. This fundamental problem captures both sensing and communication scenarios, by either (left) inferring the configuration that resulted in a measured field or (right) manipulating the configuration to achieve a desired the field.}
    \label{fig:general_opt}
\end{figure}

We next explicitly show how we can exploit the signatures induced by diffraction-inducing elements for sensing and communication tasks. More specifically, in sensing, the unique signatures produced by diffracting elements reveal information about the geometry of objects in a scene, while in communications, diffracting elements can be configured to shape the RF field and selectively increase or suppress signal strength in specific areas. To formalize this idea, we first introduce a general optimization framework that highlights key considerations when using diffraction-based strategies. This framework, illustrated in Fig.~\ref{fig:general_opt}, anchors our discussion in Sections~\ref{sec:sensing} (sensing) and~\ref{sec:comms} (communications).

Let $\Theta$ be the set of possible configurations of diffracting elements in a region of interest. Let $S_\theta(\mathbf{r})$ for $\theta\in\Theta$, denote the field defined over either a Euclidean or angular space, when the scattering elements are in configuration $\theta$. Whether for sensing or communication, a high-level objective is to find a configuration $\theta$, i.e., a set of parameters for the diffraction-inducing elements, which maximizes some scoring function, $f(S_\theta(\mathbf{r}))$, giving rise to the following optimization problem:
\begin{maxi}
        {\theta}
        {f\left(S_\theta\left(\mathbf{r}\right)\right) - f_{\text{cmplx}}(\theta) }
        {\label{prob:general}}  
        {}
        \addConstraint{\theta \in}{\Theta,}{}
\end{maxi}
where $f_{\text{cmplx}}(\theta)$ is a regularizing term which penalizes more complex configurations. Problem~\ref{prob:general} captures both sensing and communication formulations, but depending on the context, the interpretation differs, as we discuss next.

\noindent\textbf{Sensing Problem Formulation (inverse problem)}: Given a measured field $S(\mathbf{r})$, the task is to identify the parameters of the diffraction-inducing elements (\(\theta^* \in \Theta\)) that result in a modeled field $S_\theta(\mathbf{r})$ best matching the measured field $S(\mathbf{r})$.

\noindent\textbf{Communication Problem Formulation (design problem)}: We are given a desired RF field, $S(\mathbf{r})$, and a set of diffraction-inducing elements  parameterized by a set $\Theta$. We seek to find $\theta^*\in\Theta$ which results in a generated RF field closest to the desired field on the domain of interest.

Problem~\ref{prob:general} provides a general principled way to harness diffraction for solving a variety of challenging tasks. In the sections that follow, we then delve into the details of how it can be used for sensing (Sec.~\ref{sec:sensing}) and communications (Sec.~\ref{sec:comms}). In the rest of this manuscript, we mainly focus on edge diffraction, as it serves as the most suitable starting point given its rich degrees of freedom. In Sec.~\ref{sec:future}, we then discuss how other forms of diffraction can be utilized in future work.  

\section{Diffraction for Sensing}
\label{sec:sensing}
The proliferation of RF signals for environmental sensing has gained significant traction in recent years. This interest is partly driven by the ability of RF signals to sense through occlusions and operate in poor lighting conditions, and partly by their greater privacy-preserving potential compared to cameras. As such, sub-6 GHz and mmWave frequencies have both emerged as the predominant spectral bands for sensing applications, supported by the increasing availability of low-cost commodity hardware such as WiFi or off-the-shelf mmWave devices~\cite{halperin2011tool, awr2243boostug}. A substantial portion of existing RF sensing research focuses on extracting motion information from environments to infer critical scene characteristics~\cite{korany2020multiple, wang2024passive, parsay2025gait}. However, imaging static objects via RF signals still remains highly challenging, yet crucial for numerous high-value applications including scene understanding, non-invasive structural health monitoring, archaeological excavation, and search and rescue operations. In this section, we set forth that leveraging diffraction for RF sensing can significantly advance the field across a range of applications, including both dynamic and static scenarios, with a particularly strong potential for imaging static objects and scene understanding.

\subsection{Diffraction as a New Foundation for Scene Understanding}
\label{subsec:diff_as_new_found}
Across nearly all realistic environments, certain structural elements, such as edges, corners, and material transitions, serve as consistent and physically meaningful diffraction sources. As discussed in Section~\ref{sec:diffraction_primer}, such elements create distinctive electromagnetic signatures and offer a key opportunity: rather than attempting to image every surface in full detail, one can focus on detecting these diffraction-inducing elements as a compact representation of the scene. These geometric discontinuities naturally compress complex shapes into a small number of salient features, an idea echoed in perceptual psychology, where edges and junctions form the basis of object recognition~\cite{biederman1987recognition}. Thus, we argue for exploiting edges as both a compact and physically meaningful basis for RF scene understanding, leveraging their strong electromagnetic signatures as well as inherent scene compression properties.

Traditional approaches to RF imaging, such as backward ray-tracing or receive beamforming~\cite{van1988beamforming, huang2014feasibility}, assume propagation models dominated by surface reflection or diffuse scattering.  Although these methods can be effective in simple scenarios, their inherent limitations have left RF-based imaging and scene understanding a largely open research area.

For example, a common approach is to coherently sum the received signal across the aperture to test for the presence of scatterers at specific spatial locations. However, this method often performs poorly. It also suffers from limited diversity of scattered returns on the receiver grid at lower frequencies such as WiFi (wavelengths of 5–12 cm), where most environmental surfaces act as specular reflectors rather than diffuse scatterers due to the Rayleigh roughness criterion~\cite{ogilvie1991wave}. Additionally, the computational cost of evaluating every point in the reconstruction volume scales poorly with resolution, becoming prohibitively expensive~\cite{lecci2021accuracy}.

In contrast, when transmitted waves interact with edges in the scene, they generate diffracted rays that exit along Keller cones, as discussed in Sec.~\ref{sec:diffraction_primer}, leaving clear and distinctive imprints on the receiver grid. Thus, focusing on these reliable \textbf{diffraction signatures} from edge elements offers a robust and efficient alternative to surface-based reflection models. Such an approach can then be used as stand-alone (for instance similar to the treatment of this paper) or in conjunction with surface-based modeling techniques.

\begin{figure}
    \centering
    \includegraphics[width=\linewidth]{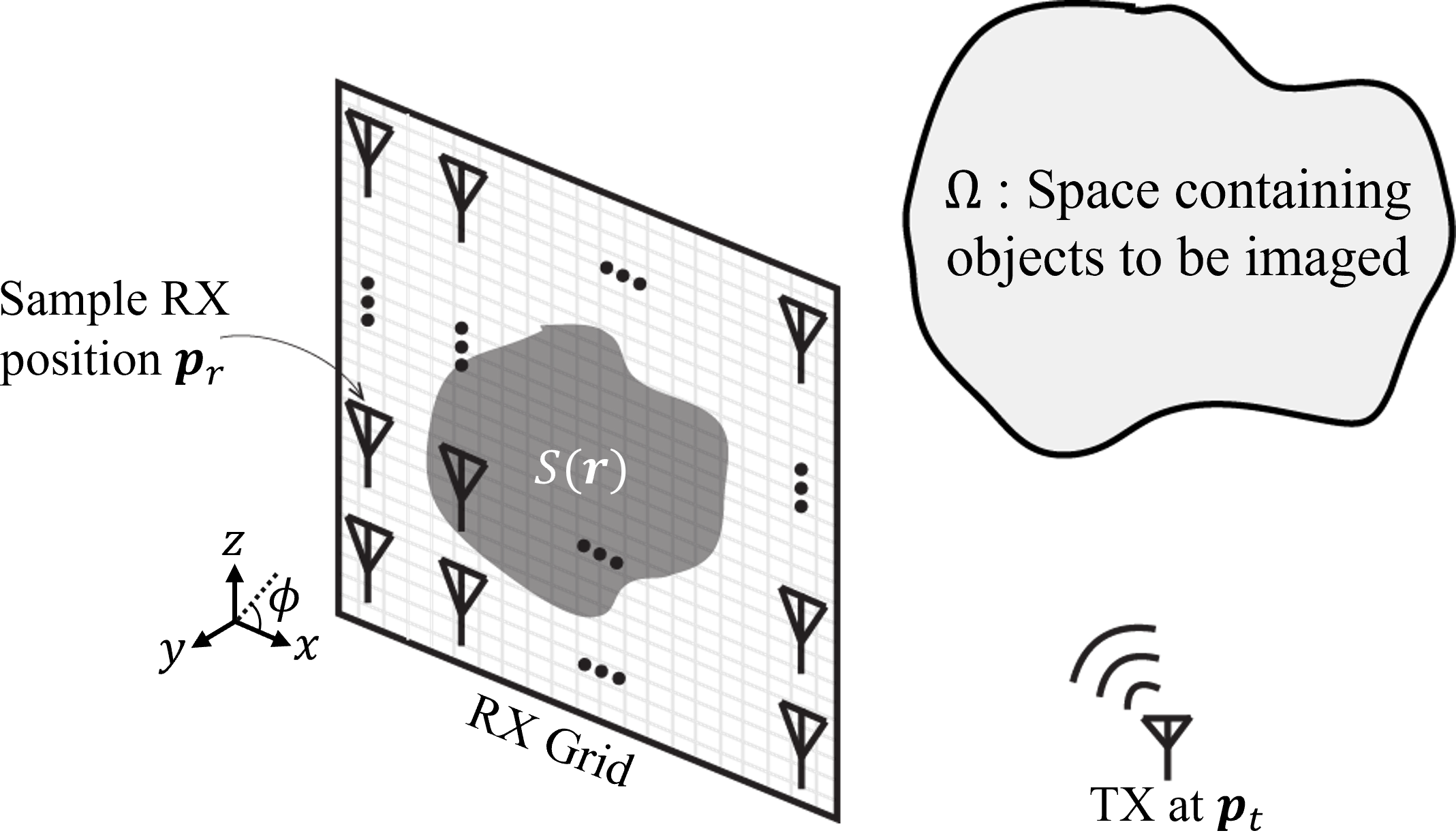}
    \caption{Schematic for a general imaging scenario: an RF transmitter emits wireless signals which scatter off of one or more extended objects in $\Omega$, subsequently captured on an extended receive aperture $RX$ as $S(\mathbf{r})\, \forall\ \mathbf{r} \in RX$. We associate each hypothesized configuration $\theta \in \Theta$ of diffracting elements in $\Omega$ with a resulting field $S_\theta(\mathbf{r})$, and seek to find the configuration $\theta^\star$ that is the most likely to have caused $S(\mathbf{r})$.}
    \label{fig:imaging_setup}
    \vspace{-0.25in}
\end{figure}

\subsection{A Framework for Diffraction-Based Sensing}
Consider the sensing setup shown in Fig.~\ref{fig:imaging_setup}, where an RF transmitter at $\mathbf{p}_t$ illuminates the scene $\Omega$. The scattered signals from extended objects within $\Omega$ are then measured across a spatially extended receiver array, denoted by $RX$. Let us define $\theta$ as the comprehensive set of parameters (spatial coordinates, edge orientation, and other relevant properties) that characterize each diffracting feature present in $\Omega$. Thus, Problem~\eqref{prob:general} represents the search for an optimal configuration of diffraction-inducing elements, denoted by $\theta^\star$, that most accurately explains the RF data $S(\mathbf{r})$ captured on $RX$. As explored in Section~\ref{sec:diffraction_primer}, each diffracting surface of the extended objects in $\Omega$ generates a distinct localized signature. In principle, for any given configuration $\theta \in \Theta$, we can predict the field $S_\theta(\mathbf{r})$ across $RX$ by composing suitable propagation models drawn from a catalog of canonical diffraction phenomena, a few of which are illustrated in Fig.~\ref{fig:modes_of_diffraction}. We then seek to find the configuration $\theta^\star$ that best explains the measured field by maximizing a similarity functional $\mathcal{L}(S_\theta, S)$: 
\begin{equation}
\label{eq:general_nfb}
\theta^\star = \arg\max_{\theta \in \Theta} \; \mathcal{L}(S_\theta, S). 
\end{equation}
This formulation yields, in principle, the most physically consistent arrangement of diffracting elements in $\Omega$ with respect to the observed signal $S(\mathbf{r})$. Thus, in the setting of Problem~\ref{prob:general}, the functional $\mathcal{L}$ plays the role of objective $f$. However, solving this optimization problem over the entire space can be computationally demanding.  Moreover, the above formulation does not explicitly take into account the interference caused by reflection or scattering off of non-diffracting elements.  We then next delve into the details on how to efficiently solve this optimization problem, with a particular attention to the cross-interference effects among diffracting elements as well as between diffracting and non-diffracting elements. Specifically, we present our recent work~\cite{pallaprolu2022wiffract, pallaprolu2023analysis} that explores the novel paradigm of using diffraction for imaging the edges of extended objects, and further showcases several real-world experiments. For clarity, we adopt the following working definition of what constitutes an edge to an incident wave:
\begin{definition}[Edge]
Let $\lambda$ denote the wavelength of the incident RF wave.  A point on a surface appears as an edge to the 
wave if its local radius of curvature $r$ satisfies $r < \lambda/2$~\cite{ross1971scattering}.
\end{definition}

We next conclude this section by examining
emerging applications of edge diffraction, including outdoor-to-indoor (O2I) localization~\cite{duggal2025diffraction} and enhanced GPS-based human sensing~\cite{dong2024gpsense}.

\subsection{Edge Diffraction for Imaging with Commodity WiFi}
We next describe the imaging pipeline we proposed in \cite{pallaprolu2022wiffract}, which leverages the Geometrical Theory of Diffraction (GTD) to efficiently characterize the signatures generated by Keller cones from the edges of extended objects when illuminated by a WiFi transmitter. As depicted in Fig.~\ref{fig:gtd_conics} and formalized in Theorem~\ref{thm:keller_law}, the signature of the Keller cone (conic section) captured on the receiver grid depends on the orientation and spatial configuration of the diffracting edge (position, azimuthal and elevation angles), as well as the relative positioning of the transmitter ($\mathbf{p}_t$) and the receiver grid ($RX$). However, due to the limited bandwidth of commodity WiFi systems,~\cite{pallaprolu2022wiffract} restricts its analysis to edges that lie within a plane parallel to the vertical receiver grid. Consequently, the edge orientation is effectively described by a single parameter $\theta = \phi$, which denotes the angle formed by the edge with respect to the positive x-axis (see Fig.~\ref{fig:imaging_setup}). Furthermore, as commodity WiFi devices do not provide reliable phase information, $S(\mathbf{r})$ represents the received power over $RX$, and we denote $\bar{S}(\mathbf{r})$ as the received power measurement after subtracting\footnote{Background subtraction is performed by measuring the scene without the object present. In cases where the object cannot be removed, dominant contributions such as the direct path and ground reflection can be estimated and subtracted~\cite{karanam2023foundation}.} the impact of background scatterers beyond $\Omega$. For a receiver located at $\mathbf{p}_r \in RX$, we define the set $\omega_{\mathbf{p}_r} \subset \Omega$, as the set of locations of all scatterers in $\Omega$ (including specular reflectors, edges, and other diffracting elements) that contribute to the field at $\mathbf{p}_r$. From~\cite[Lemma 4.1]{pallaprolu2022wiffract}, we then have:
\begin{align}\label{eq:power_new_sub}
    &\bar{S}(\mathbf{p}_r)\!\approx\!2\mathcal{R} \left\{\!\sum_{\mathbf{p}_o \in \omega_{\mathbf{p}_r}}\!\!\!\!\Lambda(\mathbf{p}_o, \mathbf{p}_t, \mathbf{p}_r) g^*(\mathbf{p}_t, \mathbf{p}_r) g(\mathbf{p}_o, \mathbf{p}_r)\!\right\},
\end{align}
where $\mathcal{R}$ is the real part of the argument, $\Lambda(\mathbf{p}_o, \mathbf{p}_t, \mathbf{p}_r)$ is the amplitude attenuation of the $\mathbf{p}_t \to \mathbf{p}_o\to \mathbf{p}_r$ path, and $g(\mathbf{p}_o, \mathbf{p}_r) = \exp(-j\frac{2\pi}{\lambda}||\mathbf{p}_o - \mathbf{p}_r||)$ is the Green function. See Appendix~\ref{sec:appendixA} for the derivation of Equation~\ref{eq:power_new_sub}.

Now let us consider a point $\mathbf{p}_m \in \Omega$. To test for the hypothesis that there exists an edge at $\mathbf{p}_m$ with orientation $\phi$,~\cite{pallaprolu2022wiffract} proposed an imaging kernel $\kappa$ that specifically selects points on $RX$ that lie on the corresponding Keller cone:
$\kappa(\mathbf{p}_m, \mathbf{p}_r, \mathbf{p}_t, \phi) =  g(\mathbf{p}_t, \mathbf{p}_r)g^*(\mathbf{p}_m, \mathbf{p}_r)\mathds{1}_{RX_{\phi, \mathbf{p}_m}}(\mathbf{p}_r),$ 
where $RX_{\phi, \mathbf{p}_m}\subset RX$ is the conic section signature of the edge at $\mathbf{p}_m$ with orientation $\phi$, and $\mathds{1}_Q(\mathbf{r})$ is the indicator function which is $1$ if $\mathbf{r}\in Q$ and zero otherwise. This kernel is then utilized to evaluate the imaging intensity $\mathcal{I}$ (as the analogue to our similarity functional $\mathcal{L}$ in Equation~\ref{eq:general_nfb}) given by:
\begin{align}\label{eq:proposed_nfb}
    \mathcal{I}(\mathbf{p}_m; \phi, \mathbf{p}_t) = \Big| \sum_{\mathbf{p}_r \in RX} \bar{S}(\mathbf{p}_r)\kappa(\mathbf{p}_m, \mathbf{p}_r,\mathbf{p}_t, \phi)\Big|.
\end{align}
The key insight in Equation~\ref{eq:proposed_nfb} is the strategic coherent summation across specifically selected receivers to achieve edge detection. To see this, we can write $\mathcal{I}(\mathbf{p}_m; \phi, \mathbf{p}_t)=|\mathcal{I}_1 + \mathcal{I}_2|$, where:
$\mathcal{I}_1(\mathbf{p}_m; \phi, \mathbf{p}_t) = \sum_{\mathbf{p}_r\in RX_{\phi, \mathbf{p}_m}} \Lambda(\mathbf{p}_m, \mathbf{p}_t, \mathbf{p}_r)$ is the signal term that constructively combines when there is an edge at $\mathbf{p}_m$ with orientation $\phi$, and
\begin{equation*}
        \mathcal{I}_2(\mathbf{p}_m; \phi, \mathbf{p}_t) = \!\!\!\!\!\!\! \sum_{\substack{\mathbf{p}_r \in RX_{\phi, \mathbf{p}_m}\\ \mathbf{p}_{o} \in \omega_{\mathbf{p}_r}, \mathbf{p}_{o}\ne \mathbf{p}_m}}\!\!\!\Lambda(\mathbf{p}_{o}, \mathbf{p}_t, \mathbf{p}_r) g(\mathbf{p}_{o}, \mathbf{p}_r)g^*(\mathbf{p}_m, \mathbf{p}_r) \nonumber
\end{equation*}
represents interference from other scatterers in the environment. The selective use of only those receivers on the Keller conic section significantly improves the signal-to-interference ratio when compared to using the entire $RX$, as interference from other scatterers (surfaces/edges) sums incoherently with high probability, resulting in lower intensity values. Conversely, when the tested edge orientation matches the actual edge orientation, the received signals sum coherently, producing a pronounced intensity peak.

Based on this theoretical foundation, the practical implementation of the edge detection algorithm requires testing multiple candidate orientations at each spatial location. Accordingly, we define $\Phi$ as the discrete set of uniformly spaced angles in the interval $[0, \pi)$, which are tested at each $\mathbf{p}_m \in \Omega$. For a given transmitter position $\mathbf{p}_t$, the estimated local edge orientation at $\mathbf{p}_m$ is subsequently given by solving the optimization problem:
\begin{align}\label{eq:edge_hyp_test}
\phi^\star(\mathbf{p}_m; \mathbf{p}_t) = \operatorname*{arg\,max}_{\phi_i \in \Phi} \mathcal{I}(\mathbf{p}_m;\phi_i,\mathbf{p}_t).
\end{align}
The orientation $\phi^\star(\mathbf{p}_m; \mathbf{p}_t)$ is selected only if the corresponding maximum intensity exceeds a predefined threshold in order to prevent false edge detections. To ensure comprehensive edge coverage, multiple transmitters can be employed, each strategically positioned to illuminate the scene from distinct vantage points. This design choice serves two purposes: first, to mitigate the near-far effect, wherein the amplitude of the received signal from proximal edges can dominate and suppress the detectability of those from more distant edges; and second, to account for configurations in which the Keller cones produced by certain edge orientations do not intersect the receive aperture, rendering those edges effectively invisible to the receiver.

\begin{figure*}
    \centering
    \includegraphics[width=\linewidth]{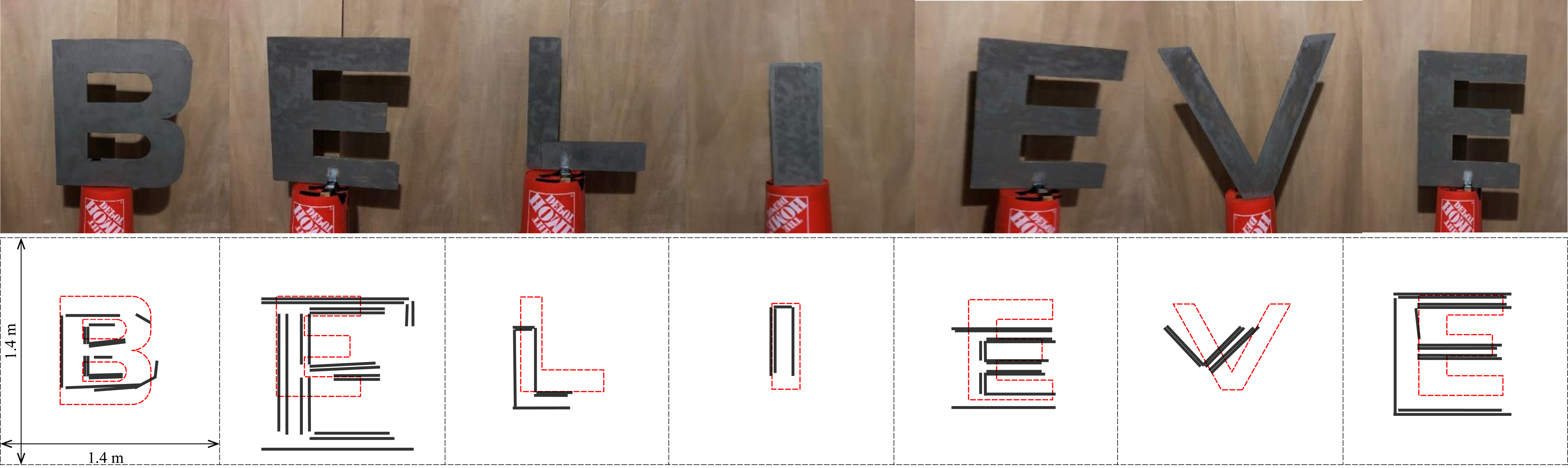}
    \caption{Commodity WiFi imaging the edges of the letters of the word ``BELIEVE" placed one by one behind a wall.  The objects are made of painted wood, with a surface paint which has a shielding attenuation that is 25dB less than aluminum. The reconstructions demonstrate the ability to recover fine structural details and character-specific shapes, even through a wall, by exploiting diffraction. See~\cite{pallaprolu2022wiffract} for more details.}  
    \label{fig:through_wall_imaging}
    \vspace{-0.2in}
\end{figure*}

The edge inferences obtained from these multiple illuminations are subsequently fused via Bayesian Information Propagation to construct a composite edge trace, or an edge image, of the extended object. The detailed implementation of this pipeline is presented in~\cite[Algorithm 1]{pallaprolu2022wiffract}.

\textbf{Remark:} The variability of conic-section signatures, along with the influence of transmitter placement, edge-receiver distance, and edge orientation, has been comprehensively analyzed in~\cite{pallaprolu2023analysis}.  Notably, it is shown in~\cite[Fig. 4]{pallaprolu2023analysis} that curved surfaces that do not visually resemble sharp edges can produce Keller cones, provided the local radius of curvature is less than half the wavelength of the incident RF field~\cite{ross1971scattering}. This implies that the proposed pipeline can not only image visibly sharp edges, but also a broader class of surfaces with a sufficiently small curvature, thereby extending the applicability of the technique to more general object geometries.
\vspace{-13pt}

\subsection{Experimental Validation}
\vspace{-2pt}
We next showcase the results of several real-world experiments for WiFi imaging by using the proposed approach, as detailed in~\cite{pallaprolu2022wiffract, pallaprolu2023analysis}. The measurement setup employs three off-the-shelf half-space panel antennas as WiFi transmitters, positioned at $\mathbf{p}_{t_1} = [3\text{m}, 0, 0]$, $\mathbf{p}_{t_2} = [0, 1.3\text{m}, 1.2\text{m}]$, and $\mathbf{p}_{t_3} = [-3\text{m}, 0, 0]$, with the origin centered at the receiver grid, which lies in the plane $y = 0$ (coordinate system shown in Fig.~\ref{fig:imaging_setup}). The receiver grid is synthesized using six vertically oriented omnidirectional antennas mounted on a scanning apparatus. This setup executes alternating horizontal scans and vertical translations in a serpentine pattern, resulting in a $42 \times 98$ grid of receivers, with vertical and horizontal spacings of $3$ cm and $1.4$ cm, respectively. Intel 5300 NICs WiFi cards on two laptops are used as receivers, and WiFi CSI power measurements are logged with CSI-Tool~\cite{halperin2011tool} at $5.32$ GHz. All experiments are conducted over an imaging plane $\Omega$ of $1.4\text{m} \times 1.4\text{m}$, located $1 - 1.5\text{m}$ from the receiver grid. See Appendix~\ref{sec:appendixB} for more details.

\begin{figure}
    \centering
    \includegraphics[width=\linewidth]{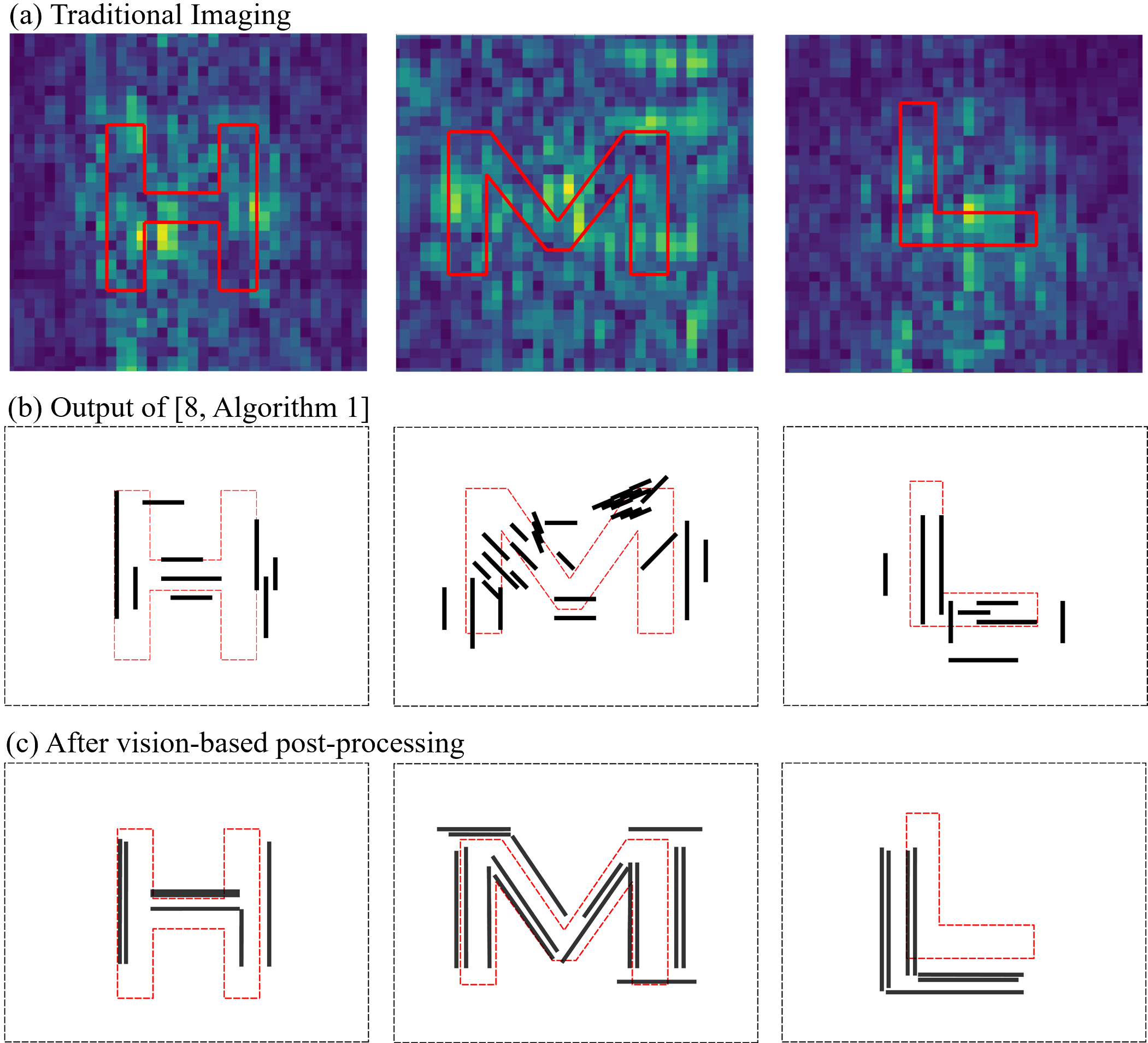}
    \vspace{-10pt}
    \caption{Imaging results for uppercase letters in a line-of-sight setting, using (a) traditional beamforming, (b)~\cite[Algorithm 1]{pallaprolu2022wiffract}, and (c) subsequent vision-based post-processing. We see that~\cite[Algorithm 1]{pallaprolu2022wiffract} captures fine structural details directly, with the post-processing serving as a refinement step that sharpens and completes the inferred shapes.}
    \label{fig:sota_comp_imaging}
    \vspace{-0.25in}
\end{figure}

To test with different edge orientations, we next show imaging sample letters of the English alphabet with WiFi. Specifically, Fig.~\ref{fig:through_wall_imaging} shows imaging results for uppercase wooden letters of the word ``BELIEVE", with dimensions $0.7\text{m}\times 0.7\text{m}$, placed one at a time behind the wall depicted in~\cite[Area 3, Fig. 7]{pallaprolu2022wiffract} (see also Fig.~\ref{fig:exp_testbeds} (a)). The wooden letters are coated with a surface paint which has a shielding attenuation that is 25dB less than aluminum. Our reconstructions incorporate a vision-based post-processing module that refines the edge images produced by~\cite[Algorithm 1]{pallaprolu2022wiffract}. More specifically, the diffraction-based images are passed through an English letter vision classifier, which serves two purposes. First, the classification accuracy provides a quantitative measure of how faithfully the diffraction-based edge images represent the true underlying structures. In our experiments, the classifier correctly classifies the letter in
26 out of 30 cases, yielding an accuracy of 86.7\%, which is substantially higher than the 3.8\% accuracy associated with random guessing. Second, the classification output is then used to refine the image by reinforcing structurally consistent edges. For example, if the classifier predicts the letter ``A", the most similar ``A" in the font dataset is used to refine the existing edges in the image. This approach is general and can be applied to other object classes, as part of future work. Further details are provided in~\cite[Sec. 5.3]{pallaprolu2022wiffract}.Fig.~\ref{fig:sota_comp_imaging} compares traditional beamforming results with the raw outputs of \cite[Algorithm 1]{pallaprolu2022wiffract} and the corresponding post‐processed reconstructions. While the conventional method produces poor results, the diffraction‐based algorithm alone distinctly traces object edges that are further refined by vision‐based post‐processing. See \cite{pallaprolu2022wiffract} for images of more letters, comparison with traditional WiFi imaging techniques and the impact of the post-processing module. 

\begin{figure}
    \centering
    \includegraphics[width=\linewidth]{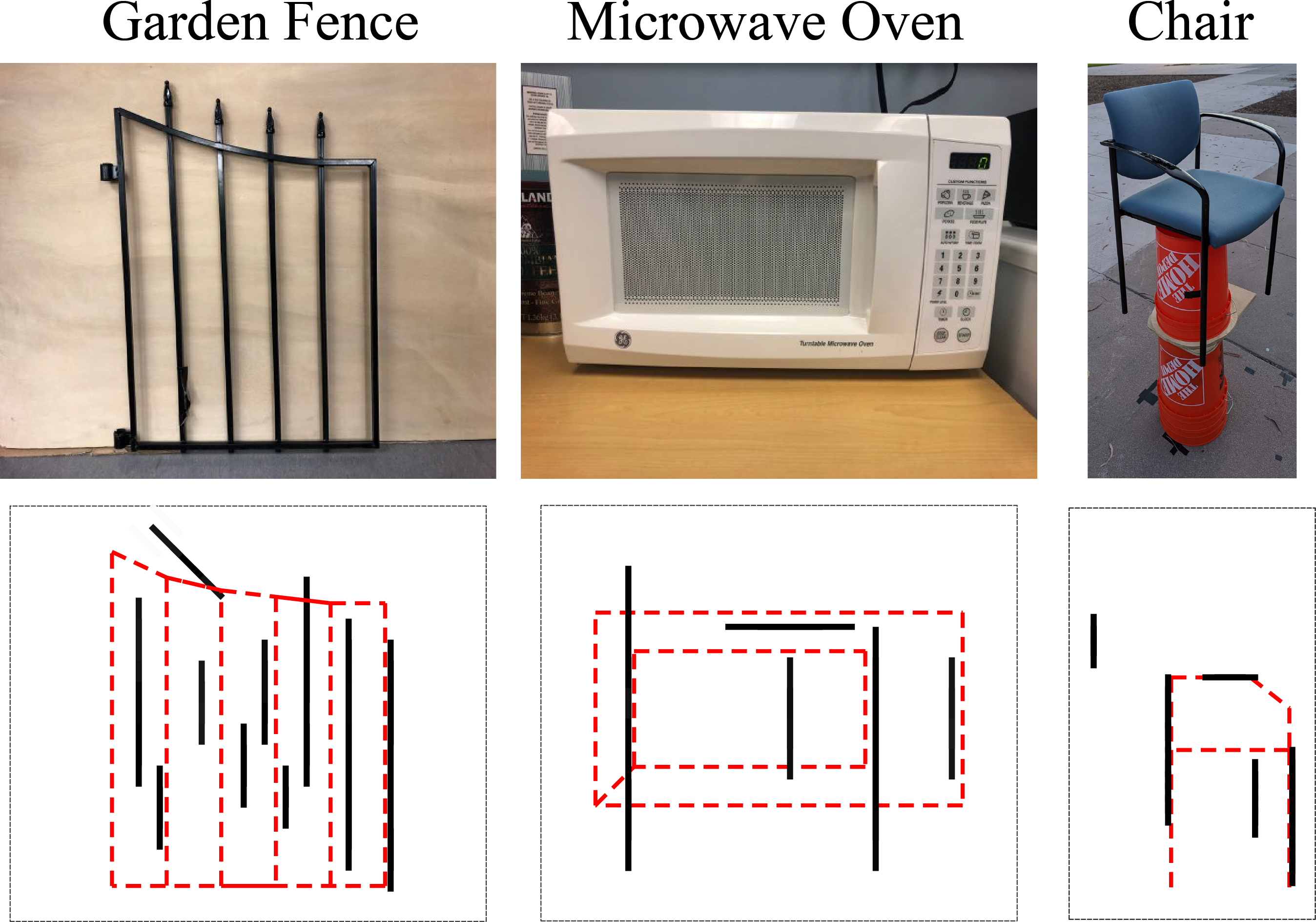}
    \vspace{-10pt}
    \caption{Imaging sample daily life objects (output of~\cite[Algorithm 1]{pallaprolu2022wiffract}), with only WiFi received power measurements and by exploiting diffraction. The objects capture a wide range of edge locations, orientations and curvatures. As part of future work, these reconstructions could be refined in post-processing by leveraging existing vision tools.}
    \label{fig:daily_life_imaging}
    \vspace{-0.2in}
\end{figure}

Fig.~\ref{fig:daily_life_imaging} then presents the output of~\cite[Algorithm 1]{pallaprolu2022wiffract} directly, when imaging a number of daily life objects featuring edges of varying radii of curvature, without any vision-based post-processing~\cite{pallaprolu2023analysis}. 
These results support the proposed hypothesis of this paper that diffraction can play a key role in the RF sensing landscape by providing rich electromagnetic signatures. 
They also highlight that edges, as compressed yet meaningful representations of objects, can be effectively leveraged for sensing. Overall, these results represent promising first steps, and future work can build upon them to further enhance the imaging quality. Sample promising future directions include: (i) optimizing transmitter placement to capture diverse edge orientations, (ii) incorporating synthetic RF data for classifier training~\cite{cai2020teaching}, (iii) improving graph-based inference for edge completion, and (iv) advanced vision-based post-processing.

Thus far, we have provided samples of our recent results to demonstrate the feasibility and efficacy of using diffraction for RF imaging. Recent research has expanded the application of edge diffraction principles and the GTD to other intriguing sensing domains, further validating the versatility of edge diffraction for wireless sensing, as we shall see next.
\vspace{-0.2in}

\subsection{Edge Diffraction for Other Sensing Applications}
\label{sec:recent_advances_diff_sensing}
Beyond imaging applications, the principles of edge diffraction have recently been adapted to address diverse sensing challenges across multiple domains. Two notable contributions have emerged that build upon the theoretical foundations of Keller cones established in our earlier work:~\cite{dong2024gpsense} has developed a passive sensing framework leveraging GPS signal diffraction for human activity detection, while~\cite{duggal2025diffraction} has applied diffraction principles to enhance outdoor-to-indoor (O2I) positioning accuracy.

\textbf{Human-Induced Edge Diffraction for GPS Signals:}~\cite{dong2024gpsense} exploits the GTD to model amplitude and phase variations when humans interact with GPS signal paths, generating dynamic Keller cones. At the GPS L1 frequency ($\lambda \approx 19$ cm), the curvature of the human torso ($\sim6$cm in radius~\cite{gordon1989anthropometric}) or shoulder ($\sim2.5$cm in radius~\cite{sahu2020geometric}) is less than half the wavelength, permitting the body to be reasonably approximated as a diffracting edge within the GTD framework. Leveraging the stability of the angle of incidence, enabled by the long-range propagation of GPS signals,~\cite{dong2024gpsense} demonstrated that signal strength fluctuation patterns correlate with the position of the receiver relative to human trajectories, directly supporting our theoretical understanding of Keller cone interactions with receive apertures. Thus, by combining reflection and diffraction models with dynamic time warping, they achieved robust human activity recognition using ubiquitous GPS signals.

\textbf{Diffraction-Aware NLoS Localization:} On the other hand,~\cite{duggal2025diffraction} incorporated diffraction path length models into Non-Line-of-Sight (NLoS) localization, focusing on O2I applications for public safety. Their approach accounts for edge diffraction from window edges as a dominant propagation mechanism. Through theoretical analysis and ray-tracing simulations, they identify two distinct multipath components arising from the Keller cones off of upper and lower window edges, and show that their relative power can be tuned via transmit-side electric field polarization control. They then propose a non-linear least squares algorithm (D-NLS) for positioning, which outperforms traditional methods, particularly in improving vertical (z-axis) accuracy for indoor localization.

Having examined these recent innovations in sensing, we now turn to a complementary question: can diffraction be harnessed not only to interpret wireless environments, but to actively shape them? The next section explores how engineered diffracting structures can be used to synthesize desired RF fields, offering a new paradigm for communication system design rooted in spatial control of such geometries.

\vspace{-0.15in}
\section{Diffraction for Communication}\label{sec:comms}
In this section, we present the use of diffraction as a new paradigm for field programming in wireless communications. Field programming is rather a new field, and can be defined as the control and shaping of the electromagnetic field propagation in space~\cite{di2020smart}. RF field programming, including beamforming and antenna pattern synthesis, plays an important role in modern communication systems by improving the signal-to-noise ratio (SNR), reducing interference, and opening the door to new types of multiple access schemes. Traditionally, achieving a desired radiation pattern requires optimizing complex-valued weights associated with omnidirectional elements in an antenna array. However, we set forth a new approach which takes advantage of unique diffraction footprints to accomplish the same goal, with greater simplicity and control. We first motivate this problem by briefly reviewing key use cases and existing methods for RF field programming. We then discuss how diffraction can be used to realize these same objectives and refer to Problem~\ref{prob:general} to highlight important considerations for designing diffracting elements for field programming. 

\vspace{-0.2in}
\subsection{RF Field Programming for Communications}
RF field programming plays a critical role in modern communication networks by enhancing signal strength along desired links and suppressing it elsewhere. This selective shaping of the electromagnetic field provides a solution to a number of challenges. For example, at high frequencies (e.g., mmWave and THz), which are central to next-generation networks, severe path and penetration losses make beamforming essential for energy-efficient, high-throughput communication~\cite{ITU-R_M.2541, ITU-R_M.2160-0}. Furthermore, field programming can improve MIMO capacity by increasing channel diversity and enabling space-division multiple access (SDMA) techniques, which focus energy on spatially separate users~\cite{dreifuerst2023CommsMag}. Moreover, attenuating gain in certain directions reduces interference and supports physical layer security (PLS) by reducing signal exposure to unintended recipients~\cite{Sanenga2020Entropy}. These benefits are particularly important in dense networks that connect thousands of end users, including autonomous vehicles, IoT devices, and mobile devices.
 
Traditional approaches to RF field programming have largely been developed in the context of large, multi-antenna arrays. Each individual antenna radiates isotropically, but by strategically distributing the transmit power and setting phase shifts across the antennas, the overall field created by the array can exhibit a complex spatial variation. Beamweights, representing the power and phase offset at each antenna, can be found using classical methods such as Minimum Variance Distortionless Response (MVDR) or Linear Constraint Minimum Variance (LCMV)~\cite{Souden2010TSP}, which solve mathematical optimization problems to ensure that the final beampattern is consistent with desired performance criteria. 

More recently, there has been significant interest in passive metasurfaces (including intelligent reflecting surfaces (IRS) and reconfigurable intelligent surfaces (RIS)), which offer a simpler and more energy efficient alternative to traditional active antenna arrays. Although metasurfaces have been somewhat successful, they also face several technical challenges. For instance, many designs call for phase shifting elements implemented as small active components that require an on-board power source and commonly possess limited phase resolution~\cite{Kebe2025JCTA}. Furthermore, stable operation of phase shifter circuitry becomes increasingly challenging at higher frequencies~\cite{yu2011book}, often introducing intermodulation distortion from other components. The complexity of the control circuitry escalates with increasing array dimensions, introducing potential failure points and significantly elevating manufacturing costs. Other implementations achieve phase shifting without active components, but the proposed elements are often complex and necessitate sophisticated manufacturing~\cite[Fig. 1]{pallaprolu2023beg}. Furthermore, they typically require a large number of elements to achieve simultaneous multi-beam steering.

Diffraction-based electromagnetic field guidance offers several compelling advantages compared to traditional receive-and-reradiate-style metasurfaces. While most phase-based radiating elements individually exhibit an omnidirectional beam pattern, edge diffraction results in a highly structured exiting beam (see Fig.~\ref{fig:gtd_conics}), whose shape can be controlled by changing the edge orientation. This makes it particularly suitable for precise beam steering applications by leveraging the rich set of conic section signatures that are scattered off of a number of illuminated edges. In other words, the orientations of a number of diffraction-inducing edge elements can be used as control knobs to shape the resulting overall RF field. Furthermore, diffraction principles can be implemented through purely geometric structures, substantially simplifying fabrication processes and reducing both manufacturing costs and power consumption requirements. Critically, diffraction-based designs can maintain consistent performance across broader frequency ranges~\cite{duggal2025impact}, offering superior bandwidth characteristics for demanding multi-band applications that would otherwise require complex active compensation in phase-shifting implementations.

\vspace{-0.2in}
\subsection{Optimization of Diffraction for Communications}
To better understand how diffraction can be used to improve communication links, consider Fig.~\ref{fig:general_opt} (right) as well as Problem~\ref{prob:general}. In our current context, this problem can be interpreted as a design problem in which we aim to configure a set of diffracting elements to improve communication in some desired way, as captured in the objective function, $f(S_\theta(\mathbf{r}))$. 
For example, we may aim to maximize the SNR on a specific link, which could be achieved by maximizing the field strength at a certain location. Alternatively, the objective may be to minimize the field strength in the direction of known possible eavesdroppers. More generally, we may aim to match a desired RF pattern, $S(\mathbf{r})$.

The set $\Theta$ represents the possible configuration of the elements, with $S_\theta(\mathbf{r})$ as the field produced by a specific configuration. The final configuration could represent either a static object to be manufactured or a specific configuration of reconfigurable hardware.

\begin{figure}
    \centering
    \includegraphics[width=0.5\linewidth]{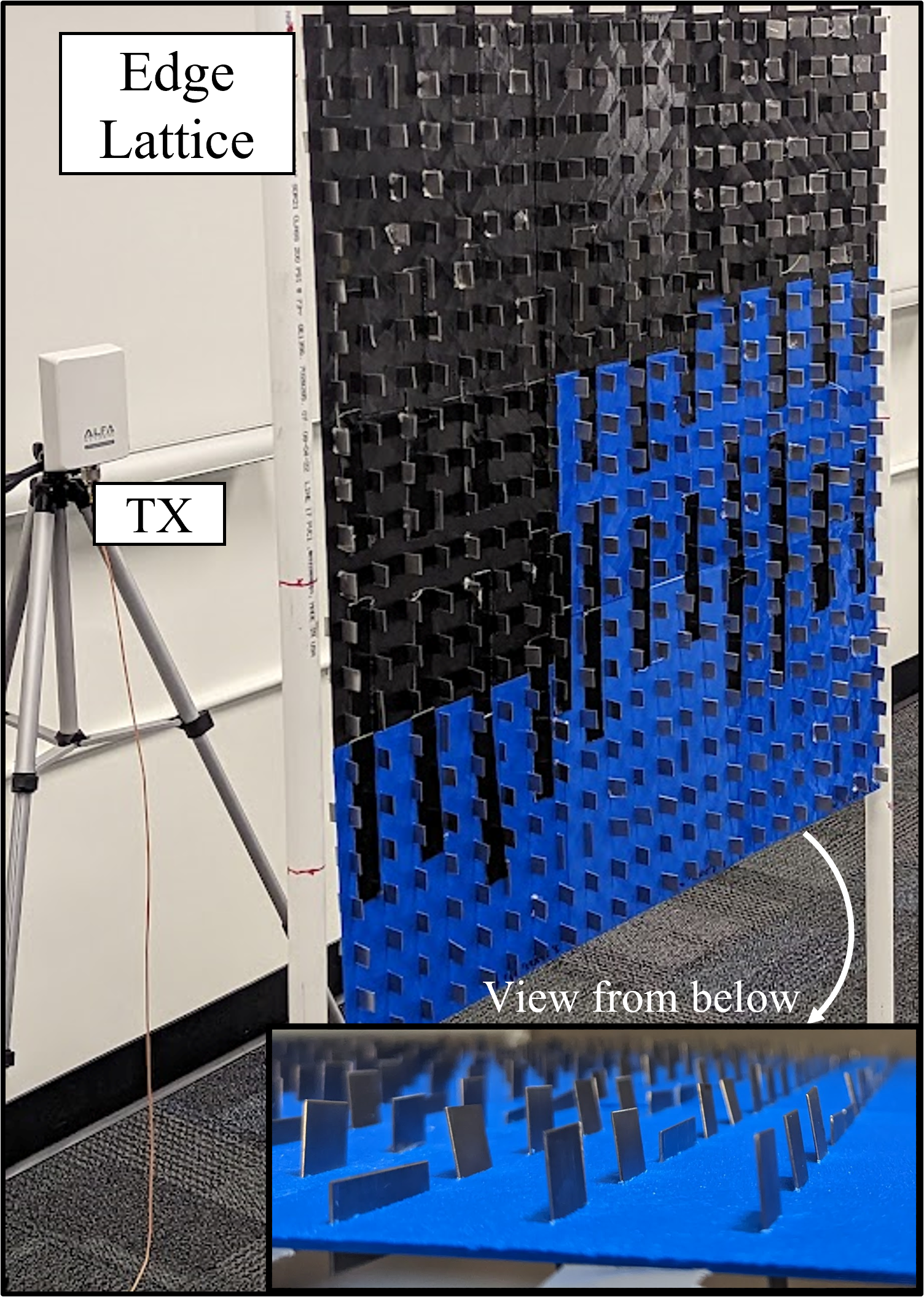}
    \caption{Diffraction-based metasurface comprised of an array of $22 \times 22$ edge elements (each edge element available at local stores, costing 7 cents) installed in front of a transmitter~\cite{pallaprolu2023beg}. The magnified section illustrates the diversity of edge orientations across the structure. Color variations in the ABS plastic are unrelated to the lattice design and result solely from available material stock. See color PDF for optimal viewing.}
    \vspace{-17pt}
    \label{fig:beg_example_metasurface}
\end{figure}

Given the non-trivial radiation mechanisms of diffraction (see Fig.~\ref{fig:modes_of_diffraction}), it is possible that multiple configurations will result in similar performance, but some of these configurations may be easier to realize than others, due to lower costs or simpler manufacturing. These considerations can be captured in the regularizer $f_{\text{cmplx}}(\theta)$, which penalizes complexity. To explore the theoretical underpinnings and practical implications, we next present a case study on multi-beam generation that demonstrates the optimization of diffracting elements for achieving desired RF field patterns.
\vspace{-20pt}

\subsection{Multi-Beam Steering utilizing Edge Diffraction}
To illustrate the possibilities of diffraction-based RF field programming, we next present our recent work from~\cite{pallaprolu2023beg}, which designs a novel diffraction-based metasurface, capable of steering up to four distinct beams simultaneously.

The ongoing development of the 6G standard has fueled recent interest in metasurfaces~\cite{katwe2024CSM, chen2025WC}, which offer a solution to a variety of challenges ranging from Non-Line-of-Sight (NLoS) communication~\cite{HUANG2025OLT} to enhanced PLS~\cite{khalid2024IoTJ}. The typical design for these metasurfaces consists of a two-dimensional array of small, closely spaced elements that work collectively to manipulate an incident RF field. Usually, this is achieved by inducing a specific phase shift at each element, similar to the approach used for active arrays~\cite{balanis2005antenna}. However, designing metasurfaces capable of producing multiple sharp beams presents a difficult challenge. State-of-the-art approaches require up to thousands of highly specialized elements and have generally been tested in only the most ideal environments, such as anechoic chambers~\cite[Fig. 1]{pallaprolu2023beg}.

\begin{figure}
    \centering
    \includegraphics[width=\linewidth]{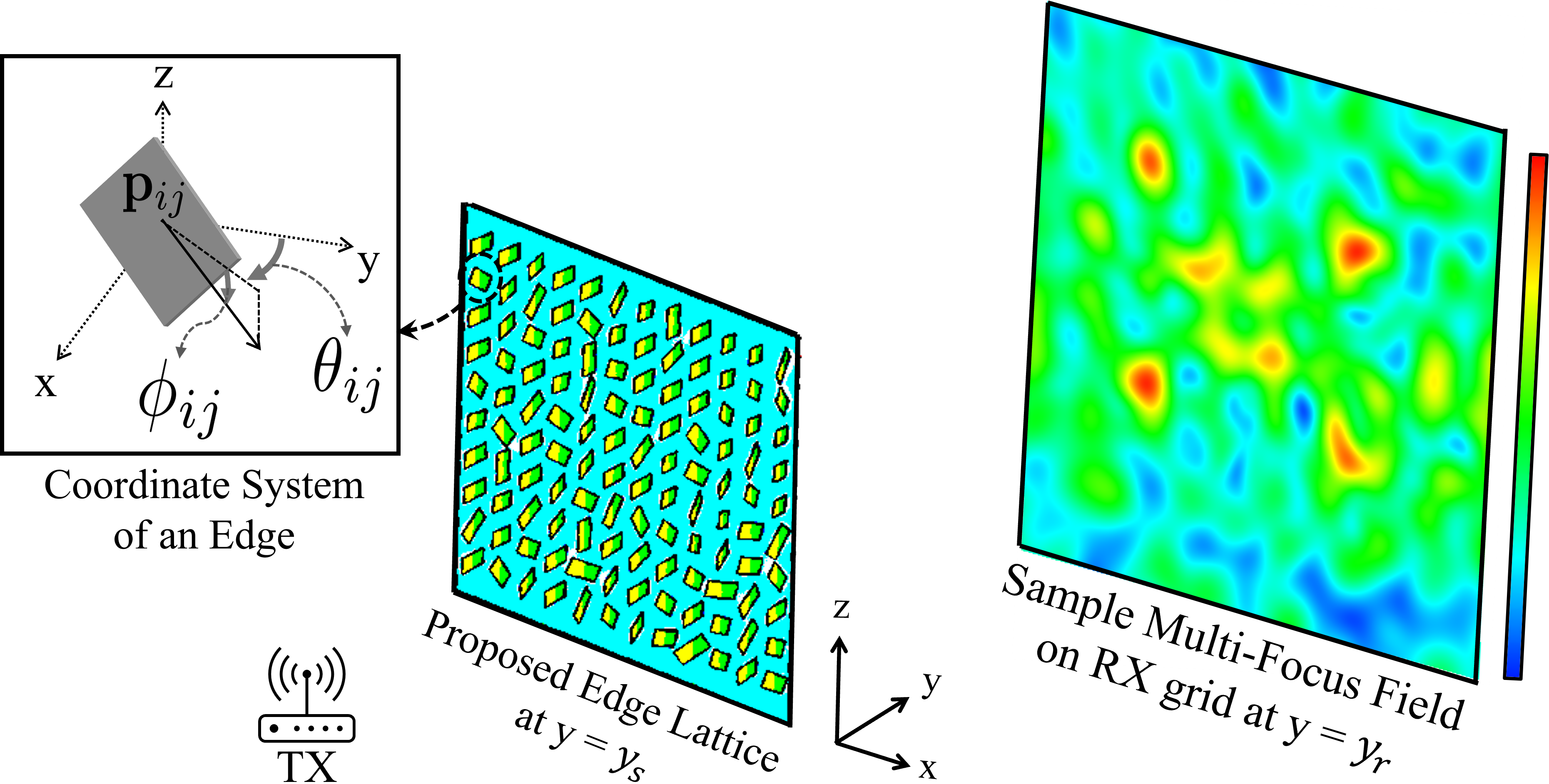}
    \caption{Schematic of a diffraction-based metasurface~\cite{pallaprolu2023beg}. The goal is to orient each edge element in a 2D array such that edge diffraction off of the elements collectively maximizes the field strength at given positions. Schematic for an individual edge element can be seen in the top left corner. See color PDF for optimal viewing.}
    \vspace{-17pt}
    \label{fig:edge_lattice_schematic}
\end{figure}

In contrast to existing approaches, utilizing diffraction can fundamentally reshape the landscape of RF field programming. To showcase its potential in this domain, we next summarize the edge-based metasurface designed in \cite{pallaprolu2023beg}. In the sequel, we use ``metasurface" to specifically refer to this design. This diffraction-based metasurface consists of a 2D array of thin, metal plates which are proportioned so that edge diffraction dominates other modes of electromagnetic interaction, and an example of such a metasurface appears in Fig.~\ref{fig:beg_example_metasurface}. By optimizing the orientations of the edges, the metasurface achieves multi-point focusing with far fewer and simpler elements. We next briefly present how the GTD informs the configuration of the edges in this metasurface to produce the desired focal points. Throughout, we use the subscript $ij$ to identify variables associated with an element located in row $i$ and column $j$.


When modeling the metasurface, the GTD gives the signature of the Keller cone associated with each edge element. Specifically, we use the following theorem~\cite[Theorem 4.1]{pallaprolu2023beg}:
\begin{theorem}
\label{thereom_bisector}
Consider an edge element located at $\mathbf{p}_{ij}$ with an edge vector of $\mathbf{e}_{ij}=[\text{cos}\phi_{ij}\text{sin}\theta_{ij}, \text{cos}\phi_{ij}\text{cos}\theta_{ij}, \text{sin}\phi_{ij}]$ (see Fig.~\ref{fig:edge_lattice_schematic} for illustration). The resulting Keller cone off of this edge element passes through location $\mathbf{p}$ if:
$$\mathbf{e}_{ij} = \frac{\mathbf{p} - \mathbf{p}_{ij}}{|\mathbf{p} - \mathbf{p}_{ij}|} + \frac{\mathbf{p}_{ij}}{|\mathbf{p}_{ij}|}.$$
\end{theorem}

While the theorem provides a constructive method for orienting each edge element such that its associated Keller cone intersects a desired focal point $\mathbf{p}$, simply steering all edge elements toward $\mathbf{p}$ may be suboptimal for focusing due to the possibility of destructive interference. It is therefore desirable to identify edge elements whose diffracted rays constructively reinforce the direct path from the transmitter to $\mathbf{p}$ and selectively involve only those elements in the focusing process. Specifically, we can assess whether the complex-valued field contribution from an edge element, $F_{\text{cone}}(\mathbf{p}, \mathbf{p}_{ij}, \mathbf{e}_{ij})$, aligns positively with the direct-path field $F_{\text{src}}(\mathbf{p}, \mathbf{x}_{\text{src}})$. Formally, the following condition is enforced:
\begin{equation}\label{eq:pos_projection}
\mathcal{R}\left\{F_{\text{cone}}^*(\mathbf{p}, \mathbf{p}_{ij}, \mathbf{e}_{ij})F_{\text{src}}(\mathbf{p}, \mathbf{x}_{\text{src}})\right\} > 0,
\end{equation}
where \(\mathcal{R}\{\cdot\}\) denotes the real part and \(^*\) the complex conjugate. Elements that do not satisfy the constructive interference condition assume an ``idle'' state, which corresponds to orienting the element vertically upward, i.e., setting $\mathbf{e}_{ij} = [0, 0, 1]$. \cite[Algorithm 1]{pallaprolu2023beg} gives the details of this approach.


\begin{figure}
    \centering
    \includegraphics[width=1\linewidth]{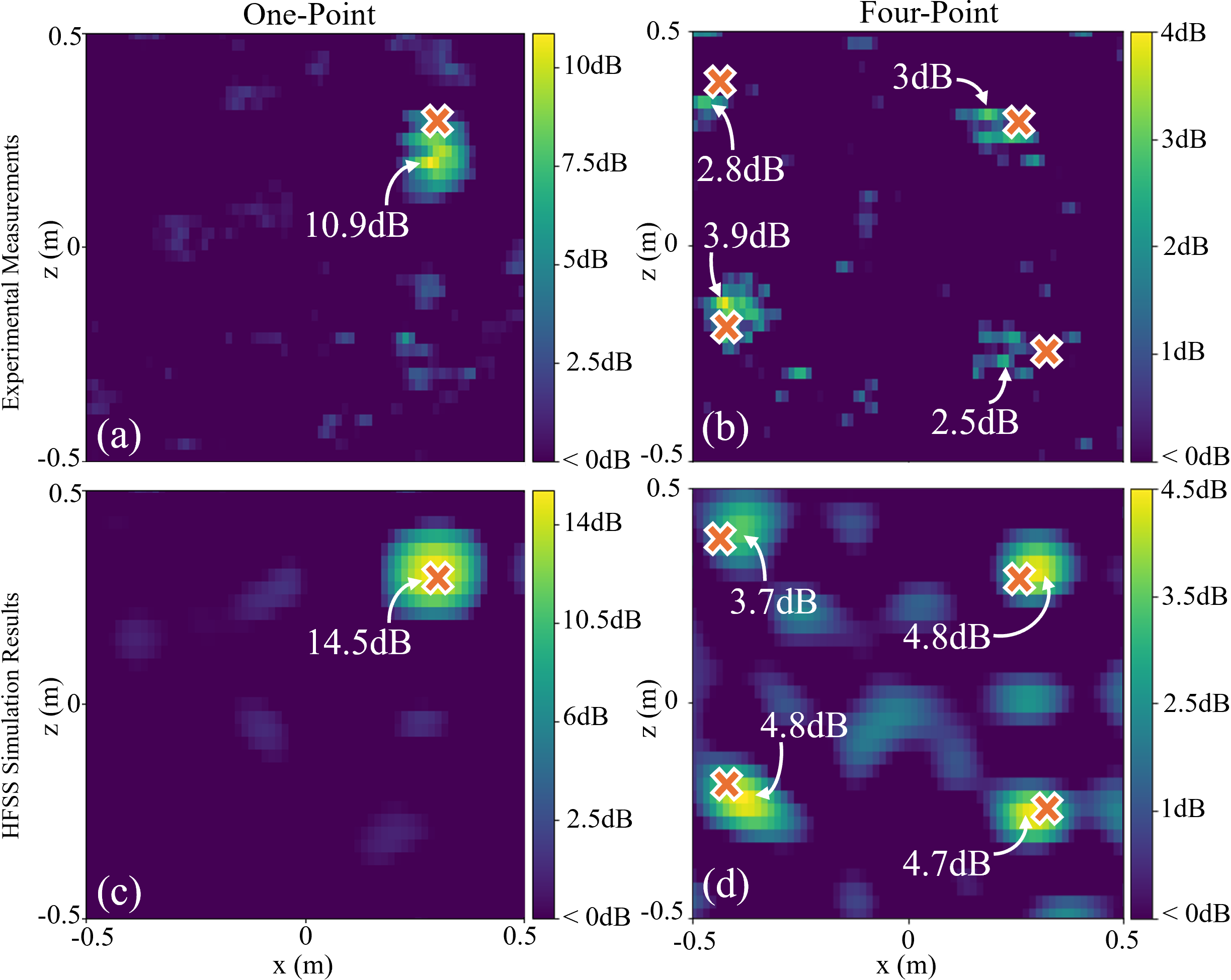}
    \vspace{-0.25in}
    \caption{Multi-beam focusing experiments using diffraction~\cite{pallaprolu2023beg} – Sample results for (a) single-point and (b) multi-point focusing in a real-world conference room. For single-point focusing, a $20\times20$ lattice of edge elements is used, while for multi-point, a $22\times22$ lattice is employed. A red “X” marks each targeted focal point, and white arrows indicate the measured peak locations. (c) and (d) show the corresponding HFSS full-wave simulation results for the same edge lattice configurations for comparison. As can be seen, the real experimental results match nicely with the HFSS full-wave simulations. See color PDF for optimal viewing.}
    \label{fig:1and4pt_focusing}
    \vspace{-0.24in}
\end{figure}

This line of reasoning can be subsequently extended to achieve multi-point focusing by dividing the edge elements into mutually disjoint subsets using \cite[Algorithm 2]{pallaprolu2023beg}, which we briefly describe next. Let $\mathcal{M} = \bigcup_{i=0}^K \mu_i$, where $\mathcal{M}$ denotes the set of all edge element indices, $\mu_i$ for $i > 0$ represents the subset assigned to the $i^{\text{th}}$ target, and $\mu_0$ corresponds to the set of idle elements. Further, let $\mathcal{T}_{ij} \subseteq \{1, \dots, K\}$ denote the set of focal points to which the edge element at $(i, j)$ can constructively contribute, as determined by Equation~\ref{eq:pos_projection}. For edge elements with $|\mathcal{T}_{ij}| > 0$, define $a_{ij} \in \mathcal{T}_{ij}$ as the target assignment for element \((i, j)\), ensuring each element contributes to at most one focal point.

Each of the focal points is equally important and therefore should receive the same number of elements. Additionally, a spatial contiguity constraint acts as a regularizing term, making partitions which spatially cluster elements assigned to the same focal point more desirable. The following optimization problem captures these considerations:
\begin{equation}
\label{eq:min_border_crossing}
\begin{aligned}
\min_{ \prod_{ij} a_{ij}\in \prod_{ij} \mathcal{T}_{ij}} \left( \frac{1}{2}\sum_{i,j} \nu_{ij} + \max_{k,l\in \{1,..., K\}}(|\mu_{k}| - |\mu_{l}|) \right),
\end{aligned}
\end{equation}
where \(\nu_{ij}\) denotes the number of neighboring elements of \((i, j)\) that belong to a different partition than \(a_{ij}\). The first term in the objective encourages spatial connectivity, while the second penalizes imbalance in the number of elements assigned to each target. Solving the problem in Equation~\ref{eq:min_border_crossing} is challenging due to its non-convex and combinatorial nature. Inspired by log-linear learning~\cite{LLLTatarenko2014} and simulated annealing, one can employ a random walk over the solution space using a Metropolis filter to find a suitable partition. 

Crucially, these analytical results are corroborated with many real-world experiments which demonstrate the ability of the diffraction-based metasurface to produce strong, highly localized and simultaneous multi-point focusing. Specifically, \cite{pallaprolu2023beg} provides the results of 13 experiments conducted in 3 distinct real-world experimental areas, including several areas with significant clutter. These experiments use a frequency of $5.18\,$GHz, and the edge elements are realized as thin $3\,$cm$\times1.5\,$cm  steel plates, costing only 7 cents each.

To manufacture the metasurface for a given set of focal points, we first partition the elements and then use Theorem~\ref{thereom_bisector} to determine the orientation for each edge in the lattice. We then 3D print a thin ($2\,$mm thick) frame made of Acrylonitrile Butadiene Styrene (ABS) with slits. These slits are angled so that when the metal plates are inserted, they are oriented appropriately. Fig.~\ref{fig:beg_example_metasurface} shows an example of such a metasurface.

The diffraction-based metasurface can produce up to 4 focal points, as shown in Fig.~\ref{fig:1and4pt_focusing}. For a single focal point, the metasurface produces $\approx11\,$dB improvement in signal strength over the field in the absence of the metasurface. For four point focusing, the metasurface results in over $3\,$dB gain, on average among the beams. To further validate the physical correctness of the design, we next compare with HFSS full-wave simulations using the same edge lattice configurations. As seen in Fig.\ref{fig:1and4pt_focusing} (c) and (d), the simulated results closely match the measured focusing behavior, despite being conducted in an idealized environment. Thus, using edge diffraction, we achieve state-of-the-art performance in signal strength improvement while using remarkably simple and inexpensive elements. Additionally, the metasurface achieves state-of-the-art (or better) performance in terms of maximum beam steering angle, beam resolution, and the number of elements required per beam, as shown in~\cite[Table 1]{pallaprolu2023beg}.

Overall, the summarized preliminary results demonstrate the feasibility of using edge-induced diffraction for RF field programming and beamforming. This highlights edge diffraction as a promising new direction for advancing future communication networks, motivating more advanced explorations for RF field programming. The next section presents a broader, forward-looking exploration of what embracing diffraction as a key element in propagation-aware system design may unlock.
\vspace{-10pt}

\section{The Road Ahead: Diffraction-Aware Systems}\label{sec:future}
Throughout this paper, we have established diffraction as a powerful and distinctive aspect of electromagnetic wave propagation, where the surface features of an extended object generate characteristic signatures that reveal crucial information about the diffracting geometry. In Sections~\ref{sec:sensing} and~\ref{sec:comms}, we presented 
preliminary investigations that strategically exploit edge diffraction phenomena, demonstrating enhanced imaging of static objects, precise O2I localization, GPS-based activity sensing, and innovative multi-point focusing using a metal-plate edge lattice as a novel metasurface. This section extends our vision further, examining how different diffraction signatures can be exploited across additional application domains. Importantly, our vision extends beyond edges to include unique signatures associated with other types of diffraction, a key dimension to be explored as part of future work.
\vspace{-10pt}
\subsection{Applications and Use Cases}
We begin by laying out promising application areas, before highlighting important challenges and potential solutions.

\textbf{Scene Understanding:} Diffraction phenomena offer a unique opportunity for RF-based scene segmentation and inference. As discussed in Sec.~\ref{sec:diffraction_primer}, diffraction produces distinct and identifiable signatures that encode geometric information far more richly than conventional specular or diffuse scattering, providing higher-dimensional electromagnetic features for scene inference. These signatures enable compact, physically meaningful descriptors: a fundamentally different representation from the pixel-wise reconstructions typical of optical systems. RF-based scene understanding has evolved considerably over the past decade, with early work initiated in~\cite{ Mostofi_TMC12}, and more recent approaches demonstrating coarse 3D reconstruction and semantic segmentation~\cite{KaranamMostofi_IPSN17, dodds2024around, chen2024rfcanvas, amatare2024rf}; however, they often depend on visual priors, precise environmental models, or dense supervised data, and their outputs lack direct physical interpretability. Diffraction-centric RF sensing defines a fundamentally different paradigm, leveraging naturally occurring geometric primitives (edges, corners, etc.) to enable more efficient, robust, and physically grounded scene understanding, without requiring complete surface reconstructions. As a future direction, the edge image outputs of daily life objects in Fig.~\ref{fig:daily_life_imaging} could be enhanced with domain-specific neural post-processing for object detection and characterization. In addition to static objects, diffraction can also be a highly useful mechanism for RF sensing and inference of dynamic, motion-induced scenes.

\textbf{Non-Line-of-Sight (NLoS) Sensing:} NLoS sensing capabilities unlock critical applications spanning structural health monitoring~\cite{zheng2021siwa, pierce2017walabot}, security surveillance~\cite{gao2024mmw}, archaeological surveying, and search-and-rescue operations~\cite{zhang2023rf}. In particular, RF-based NLoS sensing offers inherent advantages over vision, particularly in poor lighting or through occlusion. Within this space, diffraction mechanisms represent a relatively unexplored addition, offering a new modality that can help address sensing challenges. While through-wall imaging is a well-known example, NLoS scenarios more broadly include sensing around LOS blockages and other complex occlusions. As discussed in Section III-\ref{sec:recent_advances_diff_sensing}, promising recent results have demonstrated state-of-the-art performance in NLoS O2I localization by exploiting diffracted paths on Keller cones off of window edges~\cite{duggal2025diffraction}. Further development of such models, particularly those analyzing diffraction signatures from multi-layered structures or material discontinuities (as explored in acoustics~\cite{achenbach1977geometrical}) could enable breakthroughs in challenging scenarios such as the localization of surgical tools within human tissue in minimally invasive procedures~\cite{mohammadi2023uwb}.

\textbf{RF Field Programming: }
Diffraction presents an interesting opportunity to leverage the intrinsic geometric signatures of diffractive elements in the wireless ecosystem to enhance existing communication links. Our prior work~\cite{pallaprolu2023beg} demonstrated the viability of a metasurface where edge diffraction and the corresponding Keller cones played the dominant role in shaping the incident field, and future designs can benefit from integrating diffraction into established architectures. For instance, recent literature has proposed numerous theoretical models for STAR-RIS architectures that enable simultaneous transmission and reflection, but most remain at the simulation stage, with few tangible prototypes available~\cite{xu2022simultaneously}. In this context, a hybrid surface that combines diffracting and reflecting elements may offer a more tractable path toward realizing a diffraction-based STAR-RIS. Furthermore, stacked intelligent metasurfaces (SIMs)~\cite{Huang2025WCL}, which have been explored for semantic communications, could benefit from the diversity of diffraction signatures to expand control over wavefronts, enabling richer, context-aware functionalities directly at the physical layer. Lastly, the recent development of hardware architectures that perform neural computations via engineered wave propagation points to a compelling future in which computation and communication co-exist through reconfigurable, diffractive media~\cite{lin2018all}.

Beyond communication, RF field programming also holds potential for wireless power transfer (WPT), a method for remotely energizing low-power devices that has seen growing interest~\cite{Azarbahram2025TWC, hurst2025TWC}. In such scenarios, precise beamforming improves energy delivery efficiency, reducing operational costs for battery-constrained wireless sensor networks. 
\vspace{-13pt}
\subsection{Technical Challenges}
\vspace{-2pt}
The range of potential applications for exploiting diffraction in sensing and communications is broad, but realizing these capabilities introduces key challenges, which we outline next.

\textbf{Modeling and Optimization:} The success of our work in \cite{pallaprolu2022wiffract,pallaprolu2023analysis, pallaprolu2023beg} hinged on explicitly modeling edge diffraction, namely the geometry of the Keller conic section, as described by the GTD. While this approach proved effective, more detailed diffraction models could offer further accuracy and enable the inclusion of additional diffractive effects. Full-wave solvers like Ansys HFSS~\cite{ansysHFSS} provide high-fidelity results but are computationally expensive. In contrast, ray-tracing tools such as Wireless InSite~\cite{remcomInsite} and Sionna~\cite{sionnaLibrary} are scalable for large environments but lack the resolution needed for precise indoor wireless propagation. Physical-optics approximations offer a compelling middle ground~\cite{felsen1994radiation}, capturing richer wave interactions at a significantly lower cost than full-wave methods. As illustrated in Sec.~\ref{sec:diffraction_primer}, these approaches can capture generalized diffraction phenomena with significantly less computational burden than full-wave solvers, while still providing more physical accuracy than geometric ray tracing.

Machine learning presents an intriguing avenue for addressing the aforementioned challenges. In particular, data-driven approaches can model wave propagation directly as a neural network~\cite{Li2020FourierNO}, learning generalizable receiver signatures from simulated or real data and capturing complex propagation effects that are difficult to express analytically. A promising direction lies in hybrid frameworks that combine GTD with machine learning, where GTD provides a physics-based prior and ML models are used to refine or augment predictions based on observed patterns.  Such physics-informed neural networks (PINNs) have demonstrated their effectiveness in wireless propagation~\cite{zhu2024physics}, enhancing both the accuracy and efficiency of field prediction in complex environments.

\begin{figure*}
    \centering
    \includegraphics[width=\linewidth]{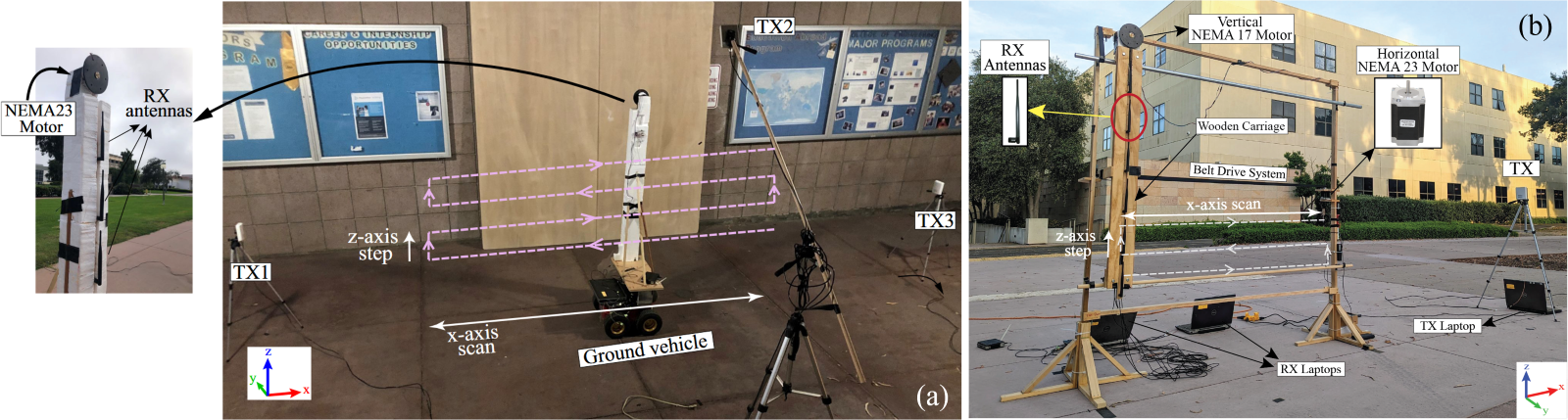}
    \vspace{-18pt}
    \caption{Experimental testbeds used for the imaging results: (a)~\cite{pallaprolu2022wiffract} uses an unmanned ground vehicle with motorized vertical antenna positioning system to synthesize a receiver grid, and (b)~\cite{pallaprolu2023analysis} implements a dual-motor wooden carriage system providing independent x-axis and z-axis control to rapidly scan an extended receiver array. In both setups, omnidirectional receive antennas and half-space panel transmit antennas are connected via SMA cables to Wi-Fi NICs of laptops, with TX and RX chains wired out from laptops to allow flexible antenna placement.}
    \label{fig:exp_testbeds}
    \vspace{-0.2in}
\end{figure*}

\textbf{Form Factors of Designed Diffracting Geometries: }Beyond modeling and optimization, realizing practical ensembles of diffracting geometries poses key challenges for advancing RF field programming. Initial explorations in passive wireless field shaping proposed optimizing conductive 3D surfaces tightly enclosing wireless access points~\cite{chan20153d,xiong2017customizing}. However, these designs often rely on coarse path-loss models, neglecting finer scattering mechanisms such as diffraction, thereby limiting the richness of achievable field patterns. Moving beyond static structures, dynamic or reconfigurable architectures represent an even greater opportunity. For instance, while our edge lattice~\cite{pallaprolu2023beg} was static, future designs could enable real-time reorientation of diffracting features through metasurface substrates based on liquid crystal elastomers~\cite{stoychev2019light}, photothermal actuators~\cite{paul2024scalable}, or shape-morphing origami materials~\cite{narumi2023inkjet}. Beyond solid-state structures, liquid and particulate media offer intriguing avenues for generating complex diffractive architectures. Inspired by classical radar countermeasures such as metallic chaff~\cite{garbacz1975advanced}, actively controlled metallic particle clouds could create strong resonant scattering environments at RF frequencies. Ferrofluid-based approaches are similarly promising: while ferro-ink solutions have been demonstrated for mmWave smart tagging~\cite{li2019ferrotag}, dynamic ferrofluid ``phase masks" could enable flexible phase modulation, akin to spatial light modulators~\cite{xie2024wavemo}, supporting selective imaging of environmental features as well as strategic RF field control. Advancing such novel form factors directly supports the broader goal of harnessing exotic geometries to unlock new levels of control over wireless environments.
\vspace{-15pt}

\section{Conclusion}\label{sec:conclusion}
This paper advocates repositioning electromagnetic diffraction from a secondary propagation effect to a primary resource for enhancing wireless systems. We argue that diffraction signatures encode valuable geometric information that can be systematically exploited. By introducing a unifying optimization framework, we demonstrate how these signatures enable novel approaches for both sensing (inferring scene geometry) and communication (shaping RF fields by configuring diffracting elements). Illustrated through recent examples in RF imaging and field programming, this diffraction-centric paradigm offers a powerful, physically grounded approach for advancing wireless capabilities. While challenges remain, particularly in modeling complex interactions, harnessing diffraction holds significant potential for developing more perceptive and robust communication systems, contributing to the goals of future wireless system design.
\vspace{-20pt}

\section*{Appendix}
\subsection{Derivation of Equation~\ref{eq:power_new_sub}}\label{sec:appendixA}
The complex baseband received signal at RX location $\mathbf{p}_r \in RX$ is a linear superposition of the direct TX path (from $\mathbf{p}_t$), background reflections (such as those from the ground or any background objects at locations $\mathbf{p}_b \in \mathcal{B}$) and object based reflections (from $\mathbf{p}_o \in \Omega$), and can be written as
$R(\mathbf{p}_r) = {\alpha}(\mathbf{p}_t, \mathbf{p}_r)g(\mathbf{p}_t,\mathbf{p}_r) +  \sum_{\mathbf{p}_b \in\mathcal{B}} \tilde{\alpha}(\mathbf{p}_b, \mathbf{p}_r)g(\mathbf{p}_b,\mathbf{p}_r) 
+ \sum_{\mathbf{p}_o \in \omega_{\mathbf{p}_r}} \tilde{\alpha}(\mathbf{p}_o, \mathbf{p}_r)g(\mathbf{p}_o,\mathbf{p}_r),
$ where $\tilde{\alpha}(\mathbf{p}_b, \mathbf{p}_r) ={\alpha}(\mathbf{p}_t, \mathbf{p}_b)  g(\mathbf{p}_t,\mathbf{p}_b){\alpha}(\mathbf{p}_b, \mathbf{p}_r)$. 

Assuming the direct path is much stronger than all scattered paths, the power of the received signal at $\mathbf{p}_r$ can be written as:
$P(\mathbf{p}_r)=C_o +  C_{\mathbf{p}_b,\mathcal{B}} + C_{\mathbf{p}_o, \omega_{\mathbf{p}_r}},$
where

$C_o = |{\alpha}(\mathbf{p}_t,\mathbf{p}_r)|^2 + \Big|\sum_{\mathbf{p}_b \in \mathcal{B}}\tilde{\alpha}(\mathbf{p}_b, \mathbf{p}_r)g(\mathbf{p}_b, \mathbf{p}_r)\Big|^2 
   +\Big|\sum_{\mathbf{p}_o \in \omega_{\mathbf{p}_r}}\tilde{\alpha}(\mathbf{p}_o, \mathbf{p}_r)g(\mathbf{p}_o, \mathbf{p}_r)\Big|^2,$ and the cross terms are
$$C_{\mathbf{p},\mathcal{P}} = 2\mathcal{R}\left\{\sum_{\mathbf{p} \in \mathcal{P}} \tilde{\alpha}(\mathbf{p}, \mathbf{p}_r){\alpha}^*(\mathbf{p}_t, \mathbf{p}_r)g(\mathbf{p}, \mathbf{p}_r)g^*(\mathbf{p}_t, \mathbf{p}_r)\right\}, \nonumber
$$
where $\mathcal{R}\{.\}$ is the real part of the argument. If power measurements are collected in the absence of the object, i.e., in the presence of only the background and the TX, we can perform a similar analysis to get:
 $   P_{bg}(\mathbf{p}_r) = |{\alpha}(\mathbf{p}_t,\mathbf{p}_r)|^2  + \Big|\sum_{\mathbf{p}_b \in \mathcal{B}}\tilde{\alpha}(\mathbf{p}_b, \mathbf{p}_r)g(\mathbf{p}_b, \mathbf{p}_r) \Big|^2 + C_{\mathbf{p}_b,\mathcal{B}}.$
By subtracting the background measurements from $P(\mathbf{p}_r)$, we obtain Equation~\ref{eq:power_new_sub}. \qed
\vspace{-10pt}

\subsection{Sample Experimental Setups Used for Imaging}\label{sec:appendixB}

Figure~\ref{fig:exp_testbeds} shows sample experimental apparatus used for imaging part in the literature. More specifically,~\cite{pallaprolu2022wiffract} uses an unmanned ground vehicle with vertically actuated omnidirectional antennas (Figure~\ref{fig:exp_testbeds} (a)) to synthesize a 42 $\times$ 140 receiver grid over a $1.2\text{m}~\times~2\text{m}$ area. The results of Figs.~\ref{fig:through_wall_imaging},~\ref{fig:sota_comp_imaging} and~\ref{fig:daily_life_imaging}, for instance, are obtained with this setup. On the other hand,~\cite{pallaprolu2023analysis} employs a dual-motor wooden carriage system (Figure~\ref{fig:exp_testbeds} (b)) to rapidly scan a 42 $\times$ 98 receiver grid with 3 cm vertical and 1.4 cm horizontal spacing. In both setups, receive antennas are connected via SMA cables to Intel 5300 NICs on laptops for CSI logging, and three off-the-shelf half-space panel antennas are used as transmitters.
\vspace{-20pt}

\bibliographystyle{IEEEtran}
{\footnotesize
\bibliography{main}}

\begin{thebibliography}{10}
\providecommand{\url}[1]{#1}
\csname url@samestyle\endcsname
\providecommand{\newblock}{\relax}
\providecommand{\bibinfo}[2]{#2}
\providecommand{\BIBentrySTDinterwordspacing}{\spaceskip=0pt\relax}
\providecommand{\BIBentryALTinterwordstretchfactor}{4}
\providecommand{\BIBentryALTinterwordspacing}{\spaceskip=\fontdimen2\font plus
\BIBentryALTinterwordstretchfactor\fontdimen3\font minus \fontdimen4\font\relax}
\providecommand{\BIBforeignlanguage}[2]{{%
\expandafter\ifx\csname l@#1\endcsname\relax
\typeout{** WARNING: IEEEtran.bst: No hyphenation pattern has been}%
\typeout{** loaded for the language `#1'. Using the pattern for}%
\typeout{** the default language instead.}%
\else
\language=\csname l@#1\endcsname
\fi
#2}}
\providecommand{\BIBdecl}{\relax}
\BIBdecl

\bibitem{ITU-R_M.2160-0}
\BIBentryALTinterwordspacing
``{{Recommendation ITU-R M.2160-0: Framework and overall objectives of the future development of IMT for 2030 and beyond}},'' International Telecommunication Union: Radiocommunication Sector, Tech. Rep., November 2023. [Online]. Available: \url{https://www.itu.int/rec/R-REC-M.2160-0-202311-I/en}
\BIBentrySTDinterwordspacing

\bibitem{luo2024channel}
C.~Luo, A.~Tang, F.~Gao, J.~Liu, and X.~Wang, ``{Channel Modeling Framework for Both Communications and Bistatic Sensing Under 3GPP Standard},'' \emph{IEEE Journal of Selected Areas in Sensors}, 2024.

\bibitem{ansysHFSS}
\BIBentryALTinterwordspacing
{Ansys Inc.}, ``{{Ansys HFSS}}: High frequency structure simulator,'' 2025, accessed: 2025-04-14. [Online]. Available: \url{https://www.ansys.com/products/electronics/ansys-hfss}
\BIBentrySTDinterwordspacing

\bibitem{CSTStudioSuite}
{Dassault Systèmes}, ``{{CST Studio Suite}},'' \url{https://www.3ds.com/products-services/simulia/products/cst-studio-suite/}, 2024, version 2024, Dassault Systèmes.

\bibitem{zhang2024radio}
F.~Zhang, C.~Zhou, C.~Brennan, R.~Wang, Y.~Li, G.~Xia, Z.~Zhao, and Y.~Xiao, ``{A radio wave propagation modeling method based on high-precision 3-D mapping in urban scenarios},'' \emph{IEEE Transactions on Antennas and Propagation}, vol.~72, no.~3, pp. 2712--2722, 2024.

\bibitem{bertoni1999radio}
H.~L. Bertoni, \emph{{Radio propagation for modern wireless systems}}.\hskip 1em plus 0.5em minus 0.4em\relax Pearson Education, 1999.

\bibitem{wu2022wifi}
D.~Wu, Y.~Zeng, F.~Zhang, and D.~Zhang, ``{WiFi CSI-based device-free sensing: from Fresnel zone model to CSI-ratio model},'' \emph{CCF Transactions on Pervasive Computing and Interaction}, pp. 1--15, 2022.

\bibitem{pallaprolu2022wiffract}
A.~Pallaprolu, B.~Korany, and Y.~Mostofi, ``{{Wiffract: a new foundation for RF imaging via edge tracing}},'' in \emph{Proceedings of the 28th annual international conference on mobile computing and networking}, 2022, pp. 255--267.

\bibitem{pallaprolu2023analysis}
------, ``{{Analysis of Keller cones for RF imaging}},'' in \emph{2023 IEEE Radar Conference (RadarConf23)}.\hskip 1em plus 0.5em minus 0.4em\relax IEEE, 2023, pp. 1--6.

\bibitem{pallaprolu2023beg}
A.~Pallaprolu, W.~Hurst, S.~Paul, and Y.~Mostofi, ``{{I beg to diffract: RF field programming with edges}},'' in \emph{Proceedings of the 29th Annual International Conference on Mobile Computing and Networking}, 2023, pp. 1--15.

\bibitem{sommerfeld1964optics}
A.~Sommerfeld, \emph{{Optics: Lectures on Theoretical Physics, Volume IV}}.\hskip 1em plus 0.5em minus 0.4em\relax New York: Academic Press, 1964, translated from the original German by O. Laporte and P. A. Moldauer.

\bibitem{felsen1994radiation}
L.~B. Felsen and N.~Marcuvitz, \emph{{Radiation and scattering of waves}}.\hskip 1em plus 0.5em minus 0.4em\relax John Wiley \& Sons, 1994.

\bibitem{keller1962geometrical}
J.~B. Keller, ``{{Geometrical theory of diffraction}},'' \emph{Journal of the Optical Society of America}, vol.~52, no.~2, pp. 116--130, 1962.

\bibitem{kouyoumjian1974uniform}
R.~G. Kouyoumjian and P.~H. Pathak, ``{{A uniform geometrical theory of diffraction for an edge in a perfectly conducting surface}},'' \emph{Proceedings of the IEEE}, vol.~62, no.~11, pp. 1448--1461, 1974.

\bibitem{luebbers1984finite}
R.~Luebbers, ``{{Finite conductivity uniform GTD versus knife edge diffraction in prediction of propagation path loss}},'' \emph{IEEE Transactions on Antennas and Propagation}, vol.~32, no.~1, pp. 70--76, 1984.

\bibitem{keller1956diffraction}
J.~Keller, ``{Diffraction of a convex cylinder},'' \emph{IRE Transactions on Antennas and Propagation}, vol.~4, no.~3, pp. 312--321, 1956.

\bibitem{chen2013ray}
X.~Chen, S.-Y. He, D.-F. Yu, H.-C. Yin, W.-D. Hu, and G.-Q. Zhu, ``{Ray-tracing method for creeping waves on arbitrarily shaped nonuniform rational B-splines surfaces},'' \emph{Journal of the Optical Society of America A}, vol.~30, no.~4, pp. 663--670, 2013.

\bibitem{brekhovskikh1960waves}
L.~M. Brekhovskikh, \emph{{Waves in Layered Media}}.\hskip 1em plus 0.5em minus 0.4em\relax New York: Academic Press, 1980.

\bibitem{goodman1996fourier}
J.~W. Goodman, \emph{{Introduction to Fourier Optics}}, 2nd~ed.\hskip 1em plus 0.5em minus 0.4em\relax New York: McGraw-Hill, 1996.

\bibitem{balanis2005antenna}
C.~A. Balanis, \emph{{Antenna Theory: Analysis and Design}}, 3rd~ed.\hskip 1em plus 0.5em minus 0.4em\relax Hoboken, NJ: Wiley-Interscience, 2005.

\bibitem{zalipaev2021diffraction}
V.~Zalipaev, ``{Diffraction of Gaussian beam by apertures and geometrical theory of diffraction},'' in \emph{2021 Days on Diffraction (DD)}.\hskip 1em plus 0.5em minus 0.4em\relax IEEE, 2021, pp. 170--176.

\bibitem{halperin2011tool}
D.~Halperin, W.~Hu, A.~Sheth, and D.~Wetherall, ``{Tool release: Gathering 802.11 n traces with channel state information},'' \emph{ACM SIGCOMM computer communication review}, vol.~41, no.~1, pp. 53--53, 2011.

\bibitem{awr2243boostug}
\emph{\BIBforeignlanguage{English}{{{AWR2243 Evaluation Module (AWR2243BOOST) mmWave Sensing Solution}}}}, Texas Instruments, available online at \url{https://www.ti.com/lit/ug/spruit8d/spruit8d.pdf}.

\bibitem{korany2020multiple}
B.~Korany, H.~Cai, and Y.~Mostofi, ``{Multiple people identification through walls using off-the-shelf WiFi},'' \emph{IEEE Internet of Things Journal}, vol.~8, no.~8, pp. 6963--6974, 2020.

\bibitem{wang2024passive}
Z.~Wang, J.~A. Zhang, H.~Zhang, M.~Xu, and J.~Guo, ``{Passive Human Tracking With WiFi Point Clouds},'' \emph{IEEE Internet of Things Journal}, 2024.

\bibitem{parsay2025gait}
A.~Parsay, M.~Torun, P.~R. Delio, and Y.~Mostofi, ``{Gait Disorder Assessment Based on a Large-Scale Clinical Trial: WiFi vs. Video vs. Doctor's Visual Inspection},'' \emph{arXiv preprint arXiv:2502.05328}, 2025.

\bibitem{biederman1987recognition}
I.~Biederman, ``{Recognition-by-components: a theory of human image understanding.}'' \emph{Psychological review}, vol.~94, no.~2, p. 115, 1987.

\bibitem{van1988beamforming}
B.~Van~Veen and K.~Buckley, ``{Beamforming: a versatile approach to spatial filtering},'' \emph{IEEE ASSP Magazine}, vol.~5, no.~2, pp. 4--24, 1988.

\bibitem{huang2014feasibility}
D.~Huang, R.~Nandakumar, and S.~Gollakota, ``{Feasibility and limits of wi-fi imaging},'' in \emph{Proceedings of the 12th ACM conference on embedded network sensor systems}, 2014, pp. 266--279.

\bibitem{ogilvie1991wave}
J.~A. Ogilvy, \emph{{Theory of Wave Scattering from Random Rough Surfaces}}.\hskip 1em plus 0.5em minus 0.4em\relax Bristol, UK: IOP Publishing, 1991.

\bibitem{lecci2021accuracy}
M.~Lecci, P.~Testolina, M.~Polese, M.~Giordani, and M.~Zorzi, ``{Accuracy versus complexity for mmWave ray-tracing: A full stack perspective},'' \emph{IEEE Transactions on Wireless Communications}, vol.~20, no.~12, pp. 7826--7841, 2021.

\bibitem{ross1971scattering}
R.~Ross and M.~Hamid, ``{Scattering by a wedge with rounded edge},'' \emph{IEEE Transactions on Antennas and Propagation}, vol.~19, no.~4, pp. 507--516, 1971.

\bibitem{duggal2025diffraction}
G.~Duggal, R.~M. Buehrer, H.~S. Dhillon, and J.~H. Reed, ``{{Diffraction-Aided Wireless Positioning}},'' \emph{IEEE Transactions on Wireless Communications}, 2025.

\bibitem{dong2024gpsense}
H.~Dong, M.~Cui, N.~Wang, L.~Qiu, J.~Xiong, and W.~Wang, ``{GPSense: Passive Sensing with Pervasive GPS Signals},'' in \emph{Proceedings of the 30th Annual International Conference on Mobile Computing And Networking}, 2024, pp. 1000--1014.

\bibitem{karanam2023foundation}
C.~R. Karanam and Y.~Mostofi, ``{A foundation for wireless channel prediction and full ray makeup estimation using an unmanned vehicle},'' \emph{IEEE Sensors Journal}, vol.~23, no.~18, pp. 21\,452--21\,462, 2023.

\bibitem{cai2020teaching}
H.~Cai, B.~Korany, C.~R. Karanam, and Y.~Mostofi, ``{Teaching RF to sense without RF training measurements},'' \emph{Proceedings of the ACM on Interactive, Mobile, Wearable and Ubiquitous Technologies}, vol.~4, no.~4, pp. 1--22, 2020.

\bibitem{gordon1989anthropometric}
C.~C. Gordon, T.~Churchill, C.~E. Clauser, B.~Bradtmiller, J.~T. McConville \emph{et~al.}, ``{Anthropometric survey of US army personnel: methods and summary statistics 1988},'' \emph{DTIC Document}, 1989.

\bibitem{sahu2020geometric}
D.~Sahu, M.~Joshi, V.~Rathod, P.~Nathani, A.~S. Valavi, and J.~D. Jagiasi, ``{Geometric analysis of the humeral head and glenoid in the Indian population and its clinical significance},'' \emph{JSES international}, vol.~4, no.~4, pp. 992--1001, 2020.

\bibitem{di2020smart}
M.~Di~Renzo, A.~Zappone, M.~Debbah, M.-S. Alouini, C.~Yuen, J.~De~Rosny, and S.~Tretyakov, ``{Smart radio environments empowered by reconfigurable intelligent surfaces: How it works, state of research, and the road ahead},'' \emph{IEEE journal on selected areas in communications}, vol.~38, no.~11, pp. 2450--2525, 2020.

\bibitem{ITU-R_M.2541}
\BIBentryALTinterwordspacing
``{{Report ITU-R M.2541: Technical feasibility of IMT in bands above 100 GHz}},'' International Telecommunication Union: Radiocommunication Sector, Tech. Rep., May 2024. [Online]. Available: \url{https://www.itu.int/pub/R-REP-M.2541-2024}
\BIBentrySTDinterwordspacing

\bibitem{dreifuerst2023CommsMag}
R.~M. Dreifuerst and R.~W. Heath, ``{Massive MIMO in 5G: How Beamforming, Codebooks, and Feedback Enable Larger Arrays},'' \emph{IEEE Communications Magazine}, vol.~61, no.~12, pp. 18--23, 2023.

\bibitem{Sanenga2020Entropy}
\BIBentryALTinterwordspacing
A.~Sanenga, G.~A. Mapunda, T.~M.~L. Jacob, L.~Marata, B.~Basutli, and J.~M. Chuma, ``An overview of key technologies in physical layer security,'' \emph{Entropy}, vol.~22, no.~11, 2020. [Online]. Available: \url{https://www.mdpi.com/1099-4300/22/11/1261}
\BIBentrySTDinterwordspacing

\bibitem{Souden2010TSP}
M.~Souden, J.~Benesty, and S.~Affes, ``{A Study of the LCMV and MVDR Noise Reduction Filters},'' \emph{IEEE Transactions on Signal Processing}, vol.~58, no.~9, pp. 4925--4935, 2010.

\bibitem{Kebe2025JCTA}
\BIBentryALTinterwordspacing
M.~Kebe, M.~C.~E. Yagoub, and R.~E. Amaya, ``{A Survey of Phase Shifters for Microwave Phased Array Systems},'' \emph{International Journal of Circuit Theory and Applications}, 2024. [Online]. Available: \url{https://onlinelibrary.wiley.com/doi/abs/10.1002/cta.4298}
\BIBentrySTDinterwordspacing

\bibitem{yu2011book}
Y.~Yu, P.~G. Baltus, and A.~H. van Roermund, \emph{{Integrated 60GHz RF Beamforming in CMOS}}.\hskip 1em plus 0.5em minus 0.4em\relax Springer Dordrecht, 2011.

\bibitem{duggal2025impact}
G.~Duggal, A.~M. Kumar, R.~M. Buehrer, H.~S. Dhillon, N.~Tripathi, and J.~H. Reed, ``{{Impact of Frequency on Diffraction-Aided Wireless Positioning}},'' \emph{arXiv preprint arXiv:2503.11993}, 2025.

\bibitem{katwe2024CSM}
M.~V. Katwe and et~al., ``{An Overview of Intelligent Meta-Surfaces for 6G and Beyond: Opportunities, Trends, and Challenges},'' \emph{IEEE Communications Standards Magazine}, vol.~8, no.~4, pp. 62--69, 2024.

\bibitem{chen2025WC}
W.~Chen, L.~Chen, Y.~Zhao, J.~Ren, and X.~Shen, ``{Metasurfaces Empowering 6G Communication and Sensing: Opportunities and Challenges},'' \emph{IEEE Wireless Commun.}, vol.~32, no.~1, pp. 158--164, 2025.

\bibitem{HUANG2025OLT}
\BIBentryALTinterwordspacing
X.~Huang, Y.~Wang, Y.~Wang, C.~Li, J.~Zhang, and S.~Li, ``{Design and implementation of metasurfaces for enhancing non-line-of-sight communication in mining tunnels},'' \emph{Optics \& Laser Technology}, vol. 182, p. 112130, 2025. [Online]. Available: \url{https://www.sciencedirect.com/science/article/pii/S0030399224015883}
\BIBentrySTDinterwordspacing

\bibitem{khalid2024IoTJ}
W.~Khalid, M.~A.~U. Rehman, T.~Van~Chien, Z.~Kaleem, H.~Lee, and H.~Yu, ``{Reconfigurable Intelligent Surface for Physical Layer Security in 6G-IoT: Designs, Issues, and Advances},'' \emph{IEEE Internet of Things Journal}, vol.~11, no.~2, pp. 3599--3613, 2024.

\bibitem{LLLTatarenko2014}
T.~Tatarenko, ``{{Proving convergence of log-linear learning in potential games}},'' in \emph{2014 American Control Conference}, 2014, pp. 972--977.

\bibitem{Mostofi_TMC12}
Y.~Mostofi, ``{{Cooperative Wireless-Based Obstacle/Object Mapping and See-Through Capabilities in Robotic Networks}},'' \emph{IEEE Transactions on Mobile Computing}, January 2012.

\bibitem{KaranamMostofi_IPSN17}
C.~R. Karanam and Y.~Mostofi, ``{{3D Through-Wall Imaging with Unmanned Aerial Vehicles Using WiFi}},'' in \emph{Proceedings of ACM/IEEE International Conference on Information Processing in Sensor Networks}, April 2017.

\bibitem{dodds2024around}
L.~Dodds, H.~Shanbhag, J.~Guan, S.~Gupta, and H.~Hassanieh, ``{Around the Corner mmWave Imaging in Practical Environments},'' in \emph{Proceedings of the 30th Annual International Conference on Mobile Computing and Networking}, 2024, pp. 953--967.

\bibitem{chen2024rfcanvas}
X.~Chen, Z.~Feng, K.~Sun, K.~Qian, and X.~Zhang, ``{RFcanvas: Modeling RF channel by fusing visual priors and few-shot RF measurements},'' in \emph{Proceedings of the 22nd ACM Conference on Embedded Networked Sensor Systems}, 2024, pp. 464--477.

\bibitem{amatare2024rf}
S.~Amatare, W.~Gao, A.~Kharel, R.~Shakya, M.~H. Rahman, X.~Shang, and D.~Roy, ``{RF-Vision: Object Characterization using Radio Frequency Propagation in Wireless Digital Twin},'' \emph{Available at SSRN 5136513}, 2024.

\bibitem{zheng2021siwa}
T.~Zheng, Z.~Chen, J.~Luo, L.~Ke, C.~Zhao, and Y.~Yang, ``{SiWa: See into walls via deep UWB radar},'' in \emph{Proceedings of the 27th Annual International Conference on Mobile Computing and Networking}, 2021, pp. 323--336.

\bibitem{pierce2017walabot}
\BIBentryALTinterwordspacing
A.~Pierce, ``{Walabot DIY Can See Into Walls},'' \emph{Technology Today}, vol.~8, no.~1, pp. 1--3, 2017, accessed: 2025-04-25. [Online]. Available: \url{https://www.technologytoday.us/Walabot_Can_See_Into_Walls.pdf}
\BIBentrySTDinterwordspacing

\bibitem{gao2024mmw}
X.~Gao, Y.~Luo, A.~Alansari, and Y.~Sun, ``{MMW-Carry: Enhancing Carry Object Detection through Millimeter-Wave Radar-Camera Fusion},'' \emph{IEEE Sensors Journal}, 2024.

\bibitem{zhang2023rf}
B.-B. Zhang, D.~Zhang, R.~Song, B.~Wang, Y.~Hu, and Y.~Chen, ``{RF-search: Searching unconscious victim in smoke scenes with RF-enabled drone},'' in \emph{Proceedings of the 29th Annual International Conference on Mobile Computing and Networking}, 2023, pp. 1--15.

\bibitem{achenbach1977geometrical}
J.~Achenbach and A.~Gautesen, ``{Geometrical theory of diffraction for three-D elastodynamics},'' \emph{The Journal of the Acoustical Society of America}, vol.~61, no.~2, pp. 413--421, 1977.

\bibitem{mohammadi2023uwb}
S.~Mohammadi, ``{UWB Radar Tool Tracking in Minimally Invasive Procedures},'' Master's thesis, University of Toronto (Canada), 2023.

\bibitem{xu2022simultaneously}
J.~Xu, Y.~Liu, X.~Mu, J.~T. Zhou, L.~Song, H.~V. Poor, and L.~Hanzo, ``{Simultaneously transmitting and reflecting intelligent omni-surfaces: Modeling and implementation},'' \emph{IEEE Vehicular Technology Magazine}, vol.~17, no.~2, pp. 46--54, 2022.

\bibitem{Huang2025WCL}
G.~Huang, J.~An, Z.~Yang, L.~Gan, M.~Bennis, and M.~Debbah, ``{Stacked Intelligent Metasurfaces for Task-Oriented Semantic Communications},'' \emph{IEEE Wireless Communications Letters}, vol.~14, no.~2, pp. 310--314, 2025.

\bibitem{lin2018all}
X.~Lin and et~al., ``{All-optical machine learning using diffractive deep neural networks},'' \emph{Science}, vol. 361, no. 6406, pp. 1004--1008, 2018.

\bibitem{Azarbahram2025TWC}
A.~Azarbahram, O.~L.~A. López, and M.~Latva-Aho, ``{Waveform Optimization and Beam Focusing for Near-Field Wireless Power Transfer With Dynamic Metasurface Antennas and Non-Linear Energy Harvesters},'' \emph{IEEE Transactions on Wireless Communications}, vol.~24, no.~2, pp. 1031--1045, 2025.

\bibitem{hurst2025TWC}
W.~Hurst and Y.~Mostofi, ``{Relay Incentive Mechanisms Using Wireless Power Transfer in Non-Cooperative Networks},'' \emph{accepted, IEEE Transactions on Wireless Communications (TWC)}, 2025.

\bibitem{remcomInsite}
\BIBentryALTinterwordspacing
{Remcom Inc.}, ``{{Wireless InSite}}: 3d wireless propagation prediction software,'' 2025, accessed: 2025-04-14. [Online]. Available: \url{https://www.remcom.com/wireless-insite-em-propagation-software}
\BIBentrySTDinterwordspacing

\bibitem{sionnaLibrary}
\BIBentryALTinterwordspacing
{NVIDIA Corporation}, ``{{Sionna}}: A library for end-to-end machine learning in wireless communications,'' NVIDIA, 2025. [Online]. Available: \url{https://nvlabs.github.io/sionna/}
\BIBentrySTDinterwordspacing

\bibitem{Li2020FourierNO}
\BIBentryALTinterwordspacing
Z.-Y. Li and et~al., ``{Fourier Neural Operator for Parametric Partial Differential Equations},'' \emph{ArXiv}, vol. abs/2010.08895, 2020. [Online]. Available: \url{https://api.semanticscholar.org/CorpusID:224705257}
\BIBentrySTDinterwordspacing

\bibitem{zhu2024physics}
E.~Zhu, H.~Sun, and M.~Ji, ``{Physics-informed generalizable wireless channel modeling with segmentation and deep learning: Fundamentals, methodologies, and challenges},'' \emph{IEEE Wireless Commun.}, 2024.

\bibitem{chan20153d}
J.~Chan, C.~Zheng, and X.~Zhou, ``{3D printing your wireless coverage},'' in \emph{Proceedings of the 2nd International Workshop on Hot Topics in Wireless}, 2015, pp. 1--5.

\bibitem{xiong2017customizing}
X.~Xiong, J.~Chan, E.~Yu, N.~Kumari, A.~A. Sani, C.~Zheng, and X.~Zhou, ``{Customizing indoor wireless coverage via 3D-fabricated reflectors},'' in \emph{Proceedings of the 4th ACM International Conference on Systems for Energy-Efficient Built Environments}, 2017, pp. 1--10.

\bibitem{stoychev2019light}
G.~Stoychev, A.~Kirillova, and L.~Ionov, ``{Light-responsive shape-changing polymers},'' \emph{Advanced Optical Materials}, vol.~7, no.~16, p. 1900067, 2019.

\bibitem{paul2024scalable}
S.~Paul, M.~R. Devlin, and E.~W. Hawkes, ``{A scalable, light-controlled, individually addressable, non-metal actuator array},'' in \emph{2024 IEEE International Conference on Robotics and Automation (ICRA)}.\hskip 1em plus 0.5em minus 0.4em\relax IEEE, 2024, pp. 684--690.

\bibitem{narumi2023inkjet}
K.~Narumi and et~al., ``{{Inkjet 4D print: Self-folding tessellated origami objects by inkjet UV printing}},'' \emph{ACM Transactions on Graphics (TOG)}, vol.~42, no.~4, pp. 1--13, 2023.

\bibitem{garbacz1975advanced}
R.~Garbacz and et~al., ``{Advanced radar reflector studies},'' \emph{The Ohio State University Electro Science Laboratory, Department of Electrical Engineering}, 1975.

\bibitem{li2019ferrotag}
Z.~Li and et~al., ``{Ferrotag: A paper-based mmwave-scannable tagging infrastructure},'' in \emph{Proceedings of the 17th Conference on Embedded Networked Sensor Systems}, 2019, pp. 324--337.

\bibitem{xie2024wavemo}
M.~Xie, H.~Guo, B.~Y. Feng, L.~Jin, A.~Veeraraghavan, and C.~A. Metzler, ``{Wavemo: Learning wavefront modulations to see through scattering},'' in \emph{Proceedings of the IEEE/CVF Conference on Computer Vision and Pattern Recognition}, 2024, pp. 25\,276--25\,285.

\end{thebibliography}

\end{document}